\documentclass[preprint,12pt]{elsarticle}

\usepackage{amssymb}
\usepackage{amsmath}

\usepackage[font=small,skip=5pt]{caption}
\usepackage{hyperref}
\usepackage{minted}
\usepackage{graphicx}
\usepackage{listings}
\usepackage{multirow}
\usepackage{xcolor}
\definecolor{codegreen}{rgb}{0,0.6,0}
\definecolor{codegray}{rgb}{0.5,0.5,0.5}
\definecolor{codepurple}{rgb}{0.58,0,0.82}
\definecolor{backcolour}{rgb}{0.95,0.95,0.92}
\lstdefinelanguage{radiant}{
  morekeywords={
    Process, Activity, Case:, Start, Intermediate, End, Start:, Intermediate:, End:,
    changes_from, in_range, is_changing, is_decreasing, is_equal,
    is_higher, is_higher_or_equal, is_increasing, is_lower,
    is_lower_or_equal, sensor, to, within, In, ConditionType
  },
  sensitive=true,
  morecomment=[l]{//},
  morecomment=[s]{/*}{*/},
  morestring=[b]",
  morestring=[b]',
  alsoletter={:},
}

\lstdefinestyle{radiantstyle}{
  language=radiant,
  keywordstyle=\color{codepurple}\bfseries,
  commentstyle=\color{codegreen}\itshape,
  stringstyle=\color{codepurple},
  basicstyle=\scriptsize\ttfamily,
  breakatwhitespace=true,         
  breaklines=true,                 
  captionpos=b,                    
  keepspaces=true,                 
  numbers=left,                    
  numbersep=5pt,                  
  showspaces=false,                
  showstringspaces=false,
  showtabs=false,                  
  tabsize=2,
  frame=single,
  belowcaptionskip=\smallskipamount,
  xleftmargin=0.03\textwidth,
}

\journal{Internet of Things}

\begin{document}

\begin{frontmatter}

\title{A Domain-specific Language and Architecture for Detecting Process Activities from Sensor Streams in IoT}


\hypersetup{
	pdfauthor={Ronny Seiger, Daniel Locher, Marco Kaufmann, Aaron F. Kurz},
	pdftitle={A Domain-specific Language and Architecture for Detecting Process Activities from Sensor Streams in IoT}
}

\author[inst1]{Ronny Seiger}
\ead{ronny.seiger@unisg.ch}

\author[inst1]{Daniel Locher}
\ead{daniel.locher@student.unisg.ch}

\author[inst1]{Marco Kaufmann}
\ead{marco.kaufmann@student.unisg.ch}

\author[inst1]{Aaron F. Kurz}
\ead{aaron.kurz@unisg.ch}

\affiliation[inst1]{organization={Institute of Computer Science},
            addressline={University of St.Gallen, Rosenbergstrasse 30}, 
            city={St.Gallen},
            postcode={9000}, 
            country={Switzerland}}

\begin{abstract}
Modern Internet of Things (IoT) systems are equipped with a large quantity of sensors providing real-time data about the current operations of their components, which is crucial for the systems' internal control systems and processes. However, these data are often too fine-grained to derive useful insights into the execution of the larger processes an IoT system might be part of. Process mining has developed advanced approaches for the analysis of business processes that may also be used in the context of IoT. Bringing process mining to IoT requires an event abstraction step to lift the low-level sensor data to the business process level. In this work, we aim to enable domain experts to perform this step using a newly developed domain-specific language (DSL) called Radiant. Radiant supports the specification of patterns within the sensor data that indicate the execution of higher level process activities. These patterns are translated to complex event processing (CEP) applications to be used for detecting activity executions at runtime. We propose a corresponding software architecture that enables online event abstraction from IoT sensor streams using the CEP applications. We evaluate these applications to monitor activity executions in smart manufacturing and smart healthcare. These evaluations are useful to inform the domain expert about the quality of activity detections based on the specified patterns and potential for improvement via additional or modified patterns and sensors.
\end{abstract}


\end{frontmatter}

\hyphenation{op-tical net-works semi-conduc-tor}

\section{Introduction} \label{sec:intro}
The ongoing pervasion of many areas of everyday life with software and technology facilitates the development of Internet of Things (IoT) systems. One important feature of IoT systems is their capability of sensing their operations to enable feedback loops between actuations in the physical world and their control in the cyber world, also known as \emph{Cyber-physical Systems} (CPS)~\cite{lee2008cyber}. Sensors in the IoT act as new data sources to inform about the CPS' operations, their interactions with other systems and entities--physical or virtual--and their surroundings~\cite{bauer2013iot}. The sensor data may also provide insights into the execution of (business) processes and process activities that IoT systems are involved in or part of~\cite{janiesch2020internet}. In contrast to analyzing the low-level control processes in one IoT device, we propose to leverage these data to analyze process executions in IoT at an abstract, business process-oriented level describing interactions among individual devices and entities in systems of IoT devices. Using Business Process Management (BPM) technologies in the context of IoT and CPS for modeling, executing and analyzing processes--also known as \emph{IoT-enhanced Business Processes}~\cite{torres2020modeling}--has already received a lot of attention from research and produced innovative results~\cite{janiesch2020internet,mangler2024internet}. Especially \emph{process mining} promises to provide comprehensive insights into process executions, including their discovery, conformance checking, and optimization as part of business process analysis~\cite{van2012process}. However, data from IoT sensors are in most cases not suitable to be directly used for process mining as they are too fine-grained to allow for gathering meaningful insights into process executions.

\begin{figure}
  \centering
  \includegraphics[width=0.85\linewidth]{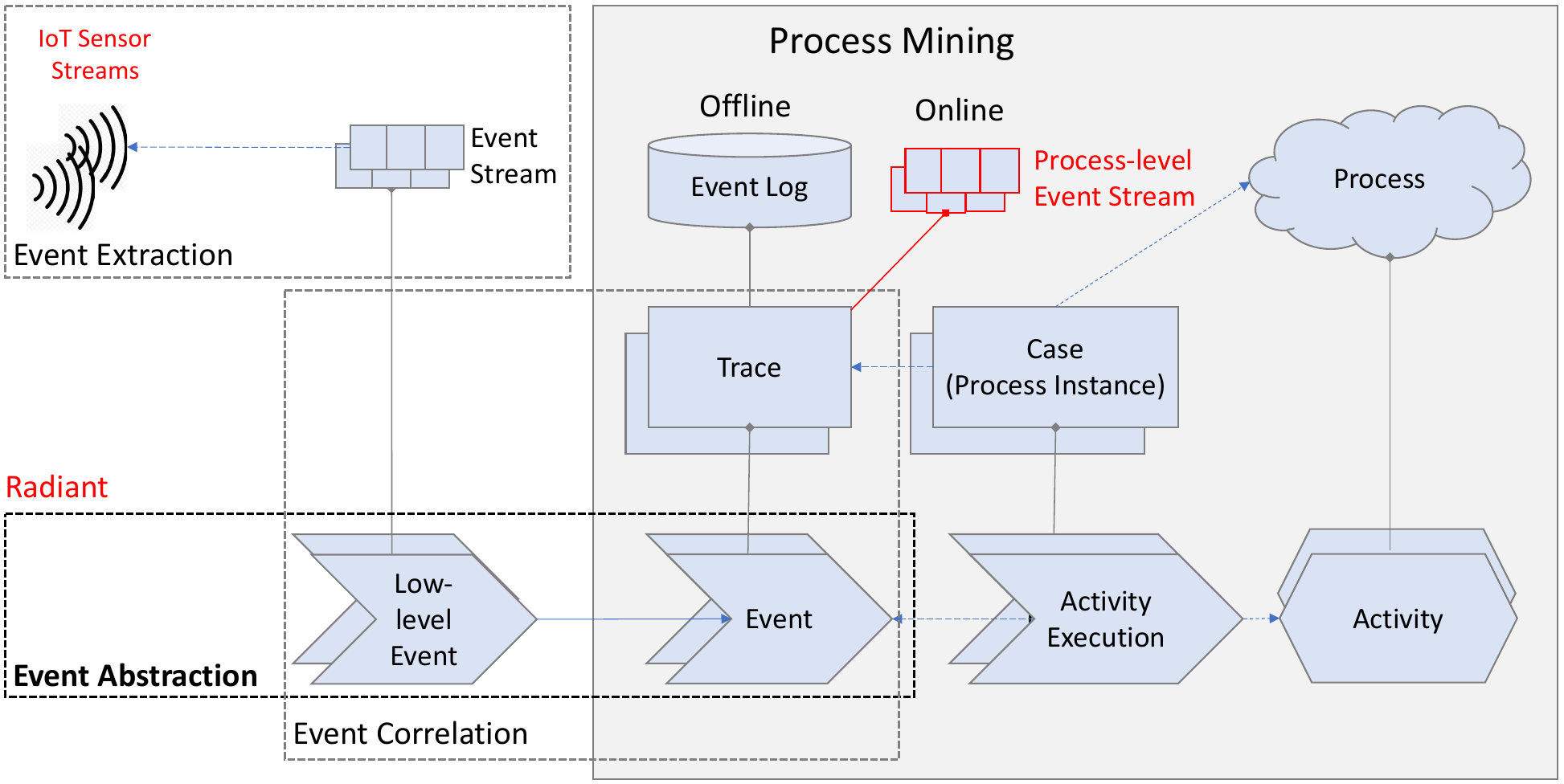}
  \caption{Radiant as means for event abstraction to transform IoT sensor streams to process-level events suitable for process mining (adapted from~\cite{diba2020extraction}).}
  \label{fig:event-abstraction}
\end{figure}

Traditional process mining assumes the existence of an \emph{event log} containing digital traces of process and activity executions which are recorded by a BPM system or process-aware information system (PAIS)~\cite{van2012process}. Events hereby indicate relevant happenings at the level of business process executions, e.g., the start or end of an activity execution. As these types of BPM systems are typically not available in IoT environments~\cite{seiger2020towards}, we propose to use the sensor data available from IoT devices to create said event logs as basis for process mining analysis. However, these sensor data are at a too fine-grained, low level informing about states of parts of the IoT system (e.g., individual motors, movements, switches, light barriers, etc.) and not necessarily about its high level operations that are more relevant for process mining. An \emph{event abstraction} step needs to be performed to \emph{lift} the sensor data to the more abstract level of process activity executions indicating \emph{what} happened and to facilitate process mining--to be able to analyze \emph{how} it happened~\cite{diba2020extraction}. The goal of our work is to enable \emph{the monitoring and analysis} of processes in IoT scenarios using process mining techniques by providing means for performing the necessary event abstraction step. Figure~\ref{fig:event-abstraction} illustrates the event abstraction problem using IoT sensors as novel data providers. Given an existing stream of IoT sensor readings, the event abstraction step transforms the IoT-based low-level events into process-level events that are associated with the execution of an activity inside a process instance. The process-level events serve as basis for process mining, either in traditional offline analysis scenarios with events stored in an event log, or in novel streaming process mining settings as highlighted in Figure~\ref{fig:event-abstraction}.

Several related approaches propose to use machine learning (ML) to perform the event abstraction step. However, these approaches need to rely on existing data to be labeled, require resources for model training, and yield incomprehensible models~\cite{wang2019deep}. Further ML-based approaches propose specialized abstractions that can only applied to specific types of sensors~\cite{cornacchia2016survey}.  In contrast to these ML-approaches, we aim to focus on the domain expert who is usually not a programmer and not versed in training machine learning models, but who possesses the relevant knowledge about changes and patterns in sensor data to perform and understand the event abstraction. This knowledge is often not available in a machine-readable form and state-of-the-art research does not provide sufficient support for domain experts to capture their knowledge about IoT sensors and sensor patterns to achieve the event abstraction. Moreover, we find a lack of execution support for existing event abstraction techniques in IoT scenarios where data is streamed from arbitrary sensors and processed at runtime, yielding process-level events.   

Guided by these observations, we formulate the first research question for this work as (\textbf{RQ1}): \emph{How can domain experts be enabled to perform process event abstraction from IoT data?} In answer to this question, we propose a novel domain-specific language (DSL) called \emph{Radiant} for event abstraction to move low-level events extracted from IoT sensors into the realm of process mining (cf.~Figure~\ref{fig:event-abstraction}). \textcolor{black}{The DSL allows IoT domain experts to specify and inspect light-weight patterns in generic IoT sensor event streams to detect process-relevant events referring to the start and end of activity executions. We provide a detailed description of Radiant's syntax with sensors, processes and activities being the most important concepts and we show how to specify patterns among sensors that lead to events related to activity executions. Our focus hereby lies on the technical feasibility of having this type of event abstraction mechanism in the form of a DSL for domain experts. We do not evaluate the DSL's usability in the scope of this article.}

Moreover, in IoT systems we see a need for event abstraction \emph{at runtime} to enable timely analysis and feedback. During process executions that also influence the physical world, it is crucial to detect the activity executions from the live IoT data streams, e.g., to react to non-conforming executions or errors detected via process mining. The second research question aims to address this requirement (\textbf{RQ2}): \emph{What is needed to detect process activity executions from abstracted process events in IoT systems at runtime?} In answer to this question, we propose a software architecture with a complex event processing (CEP) engine at its core that deploys and runs CEP applications generated from the Radiant applications written by domain experts. In contrast to related approaches on event abstraction working only in offline settings (e.g.,~\cite{janssen2020process,di2022vamos,beyel2024analyzing}) or with specific types of sensors, these CEP applications are able to process arbitrary sensor data and detect specified process-level events at runtime, which paves the way for novel online process mining and decision-making approaches~\cite{burattin2022streaming}. We showcase and evaluate the CEP applications to detect process activity executions in smart manufacturing and smart healthcare as typical IoT domains.

The contributions of this paper are as follows:
\begin{itemize}
    \item A domain-specific language (DSL) that supports the specification of generic patterns in IoT sensor data to abstract these data to the level of activities in business processes.
    \item An IDE plugin and code generator that translate specified activity detection patterns and IoT system configurations into light-weight complex event processing (CEP) applications. 
    \item The software architecture of a runtime system that supports the execution of the generated CEP applications, facilitating event-centric process mining in online and offline settings.
    \item Two case studies--including extensive datasets--demonstrating the DSL-based activity detections in smart manufacturing and smart healthcare.
\end{itemize}

The paper is structured as follows: Section~\ref{sec:preliminaries} introduces preliminaries, scenarios and background; Section~\ref{sec:related} discusses related work; Section~\ref{sec:dsl} presents the domain-specific language \emph{Radiant} and the architecture of a corresponding runtime system; Section~\ref{sec:evaluation} evaluates and discusses the activity detection; Section~\ref{sec:conclusion} concludes the paper and presents potential directions future work.

\section{Preliminaries} \label{sec:preliminaries}

In this section, we first contextualize process mining in the realms of BPM and IoT to motivate the goals and objectives of this work, and to define its scope with respect to our previous work. We then provide details on the research methodology, and we introduce two IoT scenarios from different domains that serve as running examples.

\subsection{Process Mining and IoT Data} \label{sec:pmiot}

This work is positioned as a pre-processing step in the process monitoring and analysis phases of the BPM lifecycle~\cite{dumas2013fundamentals}. This processing step preceding process mining becomes necessary due to the characteristics of IoT data representing low-level measurements--time-series data--of objects and the environment via sensors over time, which often do not match the assumed granularity of process-level events that serve as basis for process mining~\cite{zerbato2021granularity}. In the context of BPM, a \emph{process} is defined as ``a collection of inter-related events, activities, and decision points that involve a number of actors and objects, which collectively lead to an outcome that is of value to at least one customer.''~\cite{dumas2013fundamentals} We aim to monitor and detect executions of \emph{activities} (i.e.,~\emph{What} happened?) as one of the main concepts in these business processes. Activities represent atomic units of work performed by humans, software, or IoT systems in our context~\cite{dumas2013fundamentals}. The execution of an activity has a duration and it is marked by a timestamped start event and end event. These process-level activity events are the basis for analyzing process executions using event-centric process mining techniques~\cite{van2012process} (i.e.,~\emph{How} did it happen?), which we aim to enable. Abstracting IoT sensor data to activity-level events in a process is the driving theme of this work (cf.~Figure~\ref{fig:event-abstraction}). Note that we do not aim to support or enable the \emph{execution} of business processes in IoT environments that involve sensors or actuators, as shown, e.g.,~in~\cite{seiger2022integrating}, nor do we aim to develop new process mining approaches. Our goal is to make the existing corpus of process mining techniques applicable to processes in IoT settings where sensors are a novel source of event data.

In general, process mining distinguishes between 1)~\emph{discovery}, 2)~\emph{conformance checking}, and 3)~\emph{enhancement}~\cite{van2012process}. Applied to process events abstracted from IoT data, process mining allows: 1)~to discover how processes unfold in IoT-enabled environments; 2)~to check that specific process executions in and by IoT systems conform to a given, normative model of the process; and 3)~to improve processes in IoT systems based on sensor data~\cite{janiesch2020internet}. As IoT systems enable access to live sensor data streams, our goal is to perform the event abstraction at runtime (RQ2), which in turn opens up opportunities to apply and develop new process mining techniques for streaming data~\cite{burattin2022streaming}. Note that, in the context of this work, we do not aim to develop new process mining techniques for IoT data. Upon successful event abstraction, existing process mining approaches can be applied to the abstracted IoT data as shown in~\cite{seiger2022integrating}.

In previous work, we recognized novel opportunities for process mining that emerge from applying BPM technologies in IoT systems, especially to analyze process executions based on sensor data~\cite{seiger2020towards}. In~\cite{mangler2024internet} we identified several challenges that have to be addressed when using sensor data as basis for process analyses and we propose a generic framework to address these challenges. In this work we follow the framework to achieve an abstraction of IoT data to the level of process events that indicate activity executions. One goal of our work is to enable the domain expert to perform this event abstraction step (RQ1), lifting the low-level sensor data to the level of a business process, which in turn makes traditional process mining techniques accessible and applicable to IoT data. In~\cite{seiger5165943online}, we propose an alternative approach to event abstraction that relies on the domain expert to first annotate an activity execution in existing IoT data. This \emph{prototype} is then used to automatically generate and execute a service able to detect this type of activity. However, this approach suffers from incomprehensible detection logic and low detection quality when the sensor data shows a high degree of variations. We concluded that the domain expert has to be more involved in the definition of the logic and patterns used for event abstraction.

\subsection{Research Methodology and Objectives} \label{sec:dsr}

In this research, we involved experts from the two IoT domains smart manufacturing and smart healthcare in regular discussions and quarterly workshops to develop the laboratories, research objectives, and artifacts. In developing the smart healthcare setup, sensors and processes~\cite{franceschetti2023proambition}, we were supported by five medical processional from the department of infectiology at the Cantonal Hospital in St.Gallen. For the smart manufacturing setup, we discussed processes, activities, and IoT data with fellow researchers and domain experts from research departments of the \emph{Internet of Process and Things (IoPT)\footnote{\url{https://zenodo.org/communities/iopt/about}}} working group at three German universities that are familiar with these types of physical factory simulation models~\cite{malburg2020using,mangler2024internet}.
In Sections~\ref{sec:setup-manufacturing} (Smart Manufacturing) and~\ref{sec:setup-healthcare} (Smart Healthcare) we present the resulting laboratory environments that act as development and testbeds for our research. 
We followed the design science research methodology~\cite{peffers2007design} in developing the DSL (RQ1) and software architecture (RQ2) as main artifacts as follows. Thereby, we used the quarterly workshops with the domain experts to iteratively develop, refine, and evaluate the artifacts. 
\begin{itemize}
    \item Problem identification: The need for event abstraction from IoT data for process mining was derived based on our experience and insights from previous research~\cite{seiger2020towards} and existing literature~\cite{janiesch2020internet}, which elaborates on the integration of BPM and IoT technologies and their mutual benefits. In Section~\ref{sec:intro} and Section~\ref{sec:pmiot}, we introduce the two research questions~RQ1 and~RQ2, and identify the main research problem to be addressed. 
    \item Objectives of a solution: The objectives are derived from the event abstraction problem, put into the context of IoT. By studying the problem, the specific processes and IoT setups, and related approaches (cf.~Section~\ref{sec:related}), we conclude that event abstraction in IoT (functional objective) has to be (non-functional objectives):~performable and understandable by domain experts as they are the main source of knowledge regarding IoT data patterns (\emph{Objective~1});~capable of runtime data analysis to allow for timely feedback (\emph{Objective~2});~light-weight in terms of resource and data overhead for training and execution due to the potentially constraint resources available in IoT (\emph{Objective~3}); and~applicable to a wide range of sensor and IoT data in general (\emph{Objective~4})~\cite{systa2023liquidai}. We aim to address this (non-exhaustive) list of objectives within the scope of this work as they are most relevant based on our previous experience and discussions with the domain experts.
    \item Design and development: Radiant as one of the two main artifacts to address the objectives was developed following best practices in DSL design~\cite{karsai2014design,bruns2014ds}. Its syntax, the corresponding code generation, and runtime architecture are described in detail in Section~\ref{sec:dsl}. The main focus of involving the domain experts in the design of the DSL was to identify and discuss the specific patterns and conditions that an initial version of the language should support and how the available sensor data should be analyzed based on these patterns to identify activity-level events.
    \item Demonstration: We demonstrate the use of Radiant to specify the event abstraction for parts of two processes from the two application domains in Section~\ref{sec:radiant-examples}. Radiant is also available as an IDE Plugin and in an open software repository\footnote{\url{https://github.com/ics-unisg/radiant-iot-activity-dsl}} including several additional IoT examples. Furthermore, Figure~\ref{fig:dashboard} presents a screenshot from a dashboard showing the activity detections at runtime in correlation with the IoT data, which uses the prototypical implementation of the proposed software architecture as proof-of-concept demonstration.
    \item Evaluation: In Section~\ref{sec:evaluation} we evaluate Radiant's validity and feasibility to be used for event abstraction by running several applications composed with Radiant to detect the activity executions in the processes from the two IoT domains. We compare the detection from the IoT data streams with automatically or manually recorded ground truth event logs by calculating typical event detection and activity recognition metrics.
    \item Communication: This article serves as main vessel to communicate the results of our research. It is complemented by a publicly available dataset and prototype implementation of Radiant and the software architecture~\cite{seiger_2025_15773369}.
\end{itemize}

\subsection{IoT Scenario: Smart Manufacturing} \label{sec:setup-manufacturing}

\begin{figure}[h!]
  \centering
  \includegraphics[width=0.9\linewidth]{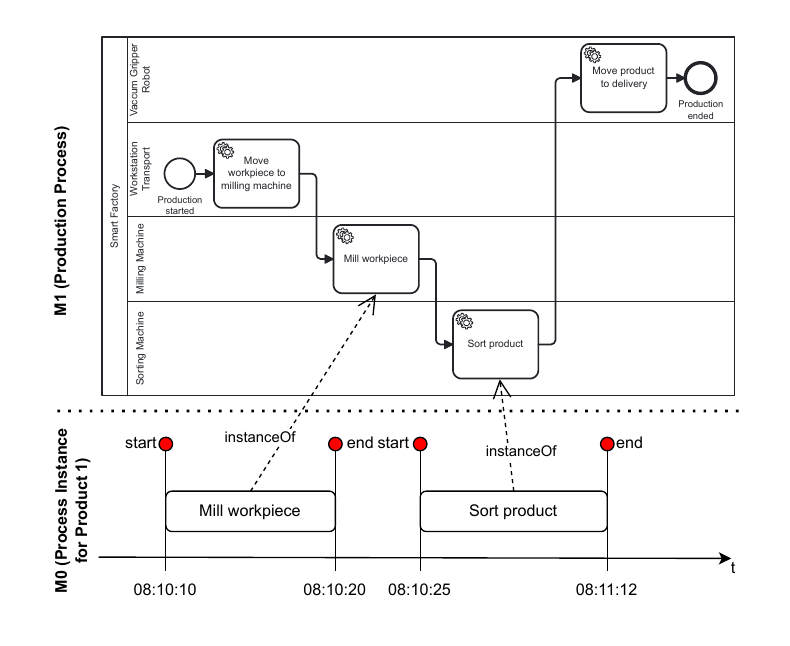}
  \caption{Top: parts of Production process model (MOF M1) in BPMN~2.0; and bottom: instances of the milling and sorting activities (MOF M0).}
  \label{fig:production-process}
\end{figure}

\paragraph{Processes} We investigate two production processes as business processes simulated in a small-scale smart factory~\cite{seiger_interactive_2023}. One process is concerned with the storage of new raw material in the factory's high-bay warehouse, which includes activities executed by a vacuum gripper robot and the warehouse. Another process implements a discrete manufacturing process with several production activities executed by a gripper robot, warehouse, oven, milling machine and sorting machine. 
To illustrate the correlations between process events, activity detection, and process models, we use abstractions for model-driven engineering introduced via the Meta-object Facility (MOF)~\cite{omg2016mof}. Figure~\ref{fig:production-process} (upper part) shows parts of the process model in BPMN~2.0~\cite{BPMN2.0} (the MOF model layer M1~\cite{omg2016mof}) specifying the normative sequence of automated activities executed by the different stations of the factory. The figure (lower part) shows the exemplary execution of one instance each of the milling activity and sorting activity (the MOF instance layer M0). The goal of our work is to detect the timestamped \emph{start} and \emph{end} events that define an activity instance in the process by processing associated sensor data. To determine the actual size of an activity, i.e.,~the level of granularity that an activity should be detected, is within the responsibility of the domain expert. This level of detail should be adequate for the process analysis that will be performed in subsequent process mining phases~\cite{zerbato2021granularity}. 

\paragraph{IoT Setup}

\begin{figure}[h]
  \centering
  \includegraphics[width=0.65\linewidth]{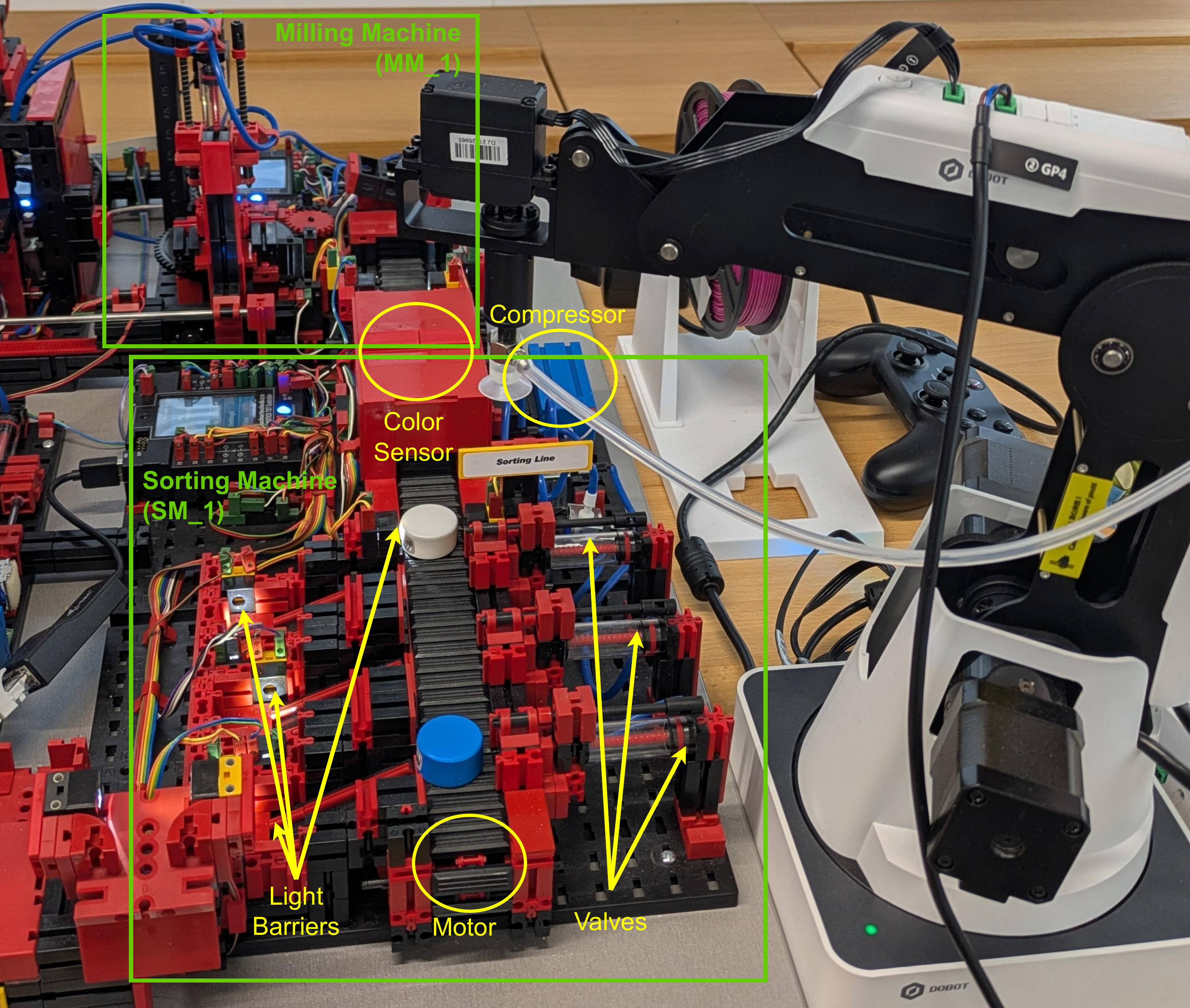}
  \caption{Parts of the sensors and actuators of the smart factory.}
  \label{fig:factory-iot}
\end{figure}

In the small-scale smart factory, there are 6 production stations with a total of 53 sensors and 24 actuators that continuously emit data about their status. This low-level time series data, which does not contain information about the process or activity executions, is the basis for the activity detection. Listing~\ref{lst:sorting} contains a timestamped data example from all sensors and actuators of the sorting machine station, which is depicted in Figure~\ref{fig:factory-iot}. Here, input sensors (\emph{sm1\_ix}) measure certain conditions (e.g., whether a light barrier is interrupted), motors (\emph{sm1\_mx}) and other output devices (\emph{sm1\_ox}) include sensors that report the status of actuators (512 is the maximum), which we also consider to be part of the IoT data. Note that this data is too fine-grained to gain any insights into the process activity that is currently being executed by this station.

\begin{lstlisting} [xleftmargin=1.8cm, linewidth=0.88\columnwidth,float,caption={Sensor data example from the sorting machine (SM\_1) in JSON.}, label={lst:sorting}]
{ "id": "b9b9969c-f86a-4d6f-950e-915b773fd363", 
 "station": "SM_1", "ts": "2023-01-30 13:06:20.27", 
 "sm1_i1_light_barrier": 0, "sm1_i2_color_sensor": 0, 
 "sm1_i3_light_barrier": 1, "sm1_i6_light_barrier": 0, 
 "sm1_i7_light_barrier": 0, "sm1_m1_speed": 512, 
 "sm1_o5_valve": 0, "sm1_o6_valve": 0, 
 "sm1_o7_valve": 512, "sm1_o8_compressor": 512 }
\end{lstlisting}

\subsection{IoT Scenario: Smart Healthcare} \label{sec:setup-healthcare}

\begin{figure}[b!]
  \centering
  \includegraphics[width=0.9\linewidth]{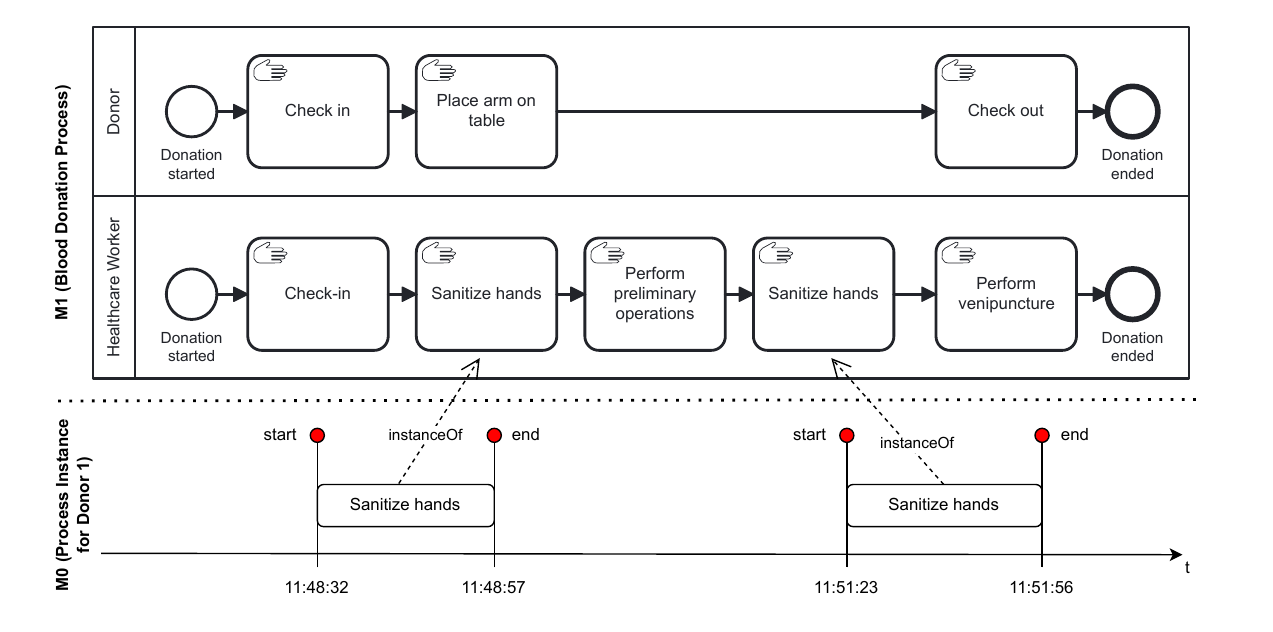}
  \caption{Top: parts of the blood donation process model (MOF M1) in BPMN~2.0; and bottom: instances of the hand sanitation activity (MOF M0).}
  \label{fig:blood-process}
\end{figure}

\paragraph{Process} We investigate the process of blood donation in a hospital setting, which includes activities performed by the healthcare worker (HCW) to prepare the instruments, disinfection steps, insertion and removal of the needle, and waste disposal~\cite{franceschetti2023proambition}. Using a combination of 20~non-intrusive sensors, we monitor the execution of the process and aim to detect the occurrence of activities in the process executed by the HCW. Note that this process is almost completely manual, i.e., executed by humans with no automation, except for the actual drawing of the blood by a machine. Figure~\ref{fig:blood-process} shows parts of the process model (MOF model layer M1) defining the normative sequence of activities in the blood donation process. The lower part shows the execution of two specific instances (MOF instance layer M0) of the activity \emph{Sanitize hands}, which are part of the same process instance, with their respective start and end events to be detected. A reasonable assumption is that \emph{one process instance} comprises the activities executed by the healthcare worker that are associated with the blood donation for \emph{one patient}.

\paragraph{IoT Setup}

\begin{figure}[b!]
  \centering
  \includegraphics[width=0.8\linewidth]{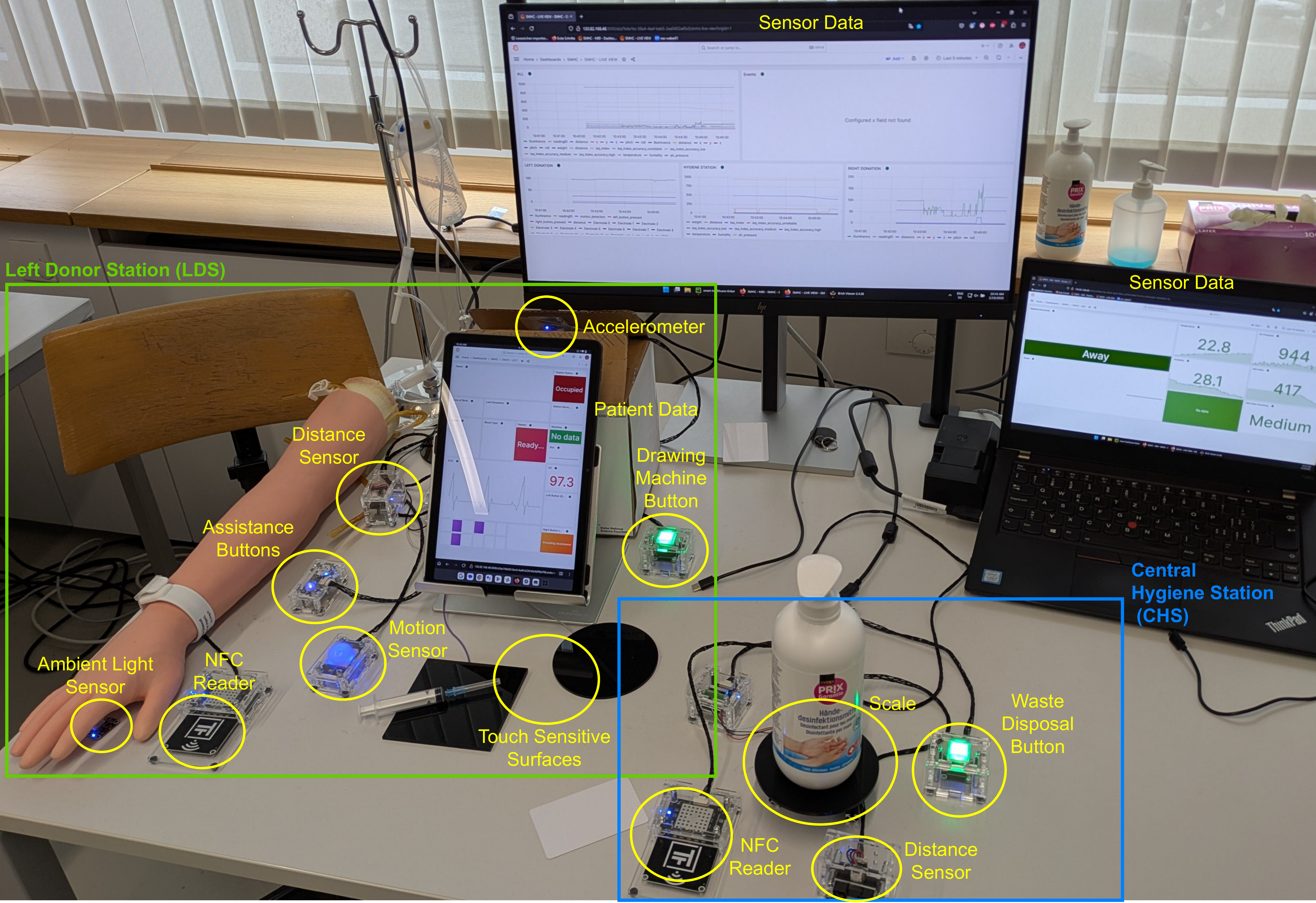}
  \caption{Sensors in smart healthcare setup.}
  \label{fig:healthcare-iot}
\end{figure}

\begin{lstlisting} [xleftmargin=1.8cm, linewidth=0.88\columnwidth,float,caption={Sensor data example from left donor station (LDS) in JSON.}, label={lst:donor}]
{ "id": "be04e317-8e5e-49ed-a05a-8b8e306ee460", 
 "station": "LDS", "ts": "2024-09-12 11:14:22.20", 
 "s25JU_ambient_light_illuminance": 446.46, 
 "s22P8_nfc_status": 0, "s22P8_nfc_tag_id": 0,
 "SjH_motion_status": 1, "VMK_button_l_pressed": 0, 
 "VMK_button_r_pressed": 0, "s23V6_button_state": 0,
 "TFf_ir_short_distance": 29.1, "VEg_accel_x": 9040, 
 "VEg_accel_y": 569, "VEg_accel_z": -3755 } 

\end{lstlisting}

To track the process activity executions in the smart healthcare setting, we deployed several rather simple, non-complex sensors--avoiding privacy-invasive sensors such as cameras--in a laboratory setup that has been developed for training and simulation purposes~\cite{franceschetti2023proambition}. Figure~\ref{fig:healthcare-iot} shows the setup including all sensors grouped into two stations--the left donor station and the central hygiene station. Together with the domain experts, we define the correlations between the values of these sensors and the executions of individual process activities. The \textit{Place arm on table} activity can for example be detected using the ambient light sensor and distance sensor on the left side of the table. The \textit{Sanitize hands} activity is detected using the scale in the central hygiene station, which increases the measured weight value significantly for a short period of time when the healthcare worker is applying pressure to the bottle containing the sanitizer. The \textit{Perform venipuncture} can be detected by analyzing the distance, motion and ambient light sensors from the left station (indicating patient movement) combined with the distance sensor from the central station (indicating HCW movement) over a short period of time. Listing~\ref{lst:donor} contains an exemplary timestamped reading of all the sensors from the left donor station in JSON format. While deploying this kind of sensor-based monitoring setup in a hospital room might be not be completely realistic, the medical experts from our partner hospital highlighted its usefulness in training settings to monitor and inform medical students and trainees about the correctness and compliance of their treatment process executions. 

\section{Related Work} \label{sec:related}

Two significant works discuss challenges and opportunities of using IoT data together with BPM systems, highlighting the issue of event abstraction from low-level data to BPM-related data~\cite{janiesch2020internet,brzychczy2025process}, and more specific challenges associated with using data from sensors in this context~\cite{mangler2024internet}. Mangler et al.~conclude that the relevance of sensors and dependencies among sensors contributing to the detection of BPM activities are difficult to be derived automatically and often need to involve domain experts. Moreover, sensor data associated with the execution of activities of the same type might be subject to variations due to external factors and parameters. Especially data from continuous sensors is affected by variations, which requires discretizations, pre-processing and abstraction steps often relying on domain expertise~\cite{mangler2024internet}. These observations confirm our focus on enabling the domain expert to define and configure the IoT-based event abstraction (cf.~RQ1).

\subsection{Event Abstraction from Sensor Data} \label{sec:rel-ml}

In~\cite{DBLP:conf/icsoc/BackmannBH0W13}, the authors discuss the monitoring of business processes by automatically generating complex event processing (CEP) queries associated with the lifecycle transitions of control flow elements. Complementary to this work,~\cite{mousheimish2016autocep} proposes to automatically learn and generate CEP rules for business activity detection from historical event traces. While our intent of monitoring the execution of business processes is similar, the authors rely on data from a BPM system to be available to trigger and log corresponding events. Assuming that BPM systems are not always available in IoT systems, we first need to perform an event abstraction step to go from the IoT data to BPM-related execution data~\cite{diba2020extraction}, which is the main functional objective of our work.

Various approaches investigate detecting and deriving BPM-related information from low-level sensor data. Janssen et al.~discuss the discovery of process models from sensor event data in~\cite{janssen2020process}, based on machine learning (ML). Automatically derived, recurring patterns in sensor data enriched with context information are used to discover human activities and habits in~\cite{di2022vamos,di2023cvamos}. In~\cite{beyel2024analyzing} the authors derive the operations of a smart car for process mining based on a statistical analysis of the car's sensor data. The authors in~\cite{rebmann2019enabling} focus on the detection of activities in manual processes based on different modalities. Large language models (LLMs) are investigated in the \emph{IoTMiner} framework by~\cite{brzychczy2025iot}. Following some pre-processing steps, the LLMs are prompted to classify segmented data from IoT based on pre-defined activity types. While these works fulfill the main functional objective of event abstraction and are applicable to generic IoT data (Objective~4), they only partially include domain expertise (e.g., in the LLM prompt), and they introduce the need for a training data being already available, potentially with additional labels (e.g., when using ML). None of them has demonstrated online capabilities as they rely on a post-mortem offline analysis for classification (Objective~3).

Even though the given problem might be suitable to be approached by ML~\cite{wang2019deep}, the training of corresponding models requires existing data and labels, is often resource-intensive and leads to large, monolithic and inexplicable models, which are hard to maintain and adjust, especially by domain experts who might not be familiar with ML (Objective~1). Many works propose trained, supervised models to detect human activities from sensors (e.g.,~wearables~\cite{GarciaCejaBCG14,cornacchia2016survey}, smartphones~\cite{yin2015human} or cameras~\cite{aggarwal2014human}). These works are only applicable to specific types of sensors, they only work in offline settings, and they require extensive existing training data and ML models that do not foster the understandability of the event abstraction (Objective~1)~\cite{nweke2019data}. Even ML-models for sensor-based activity detection that are tailored and heavily optimized for edge computing scenarios (\emph{TinyML}) are 130\,kb in size or much larger~\cite{zhou2025efficient,sharma2024efficient} and additionally require a model-serving infrastructure. We aim to develop a more general, light-weight approach that is applicable to any kind of activity--manual or automated--and arbitrary IoT sensor data (Objectives~3 and~4). Most of the ML-based approaches do not consider existing domain expertise in their models, they need to rely on data and might incorporate some expertise in the provided labels. Our goal is to treat the expertise of the domain experts regarding sensors and sensor patterns as first class citizens, without the need to have a corpus of training data already available. Moreover, many related ML-based approach are only capable of post-mortem detection and classification of activities. With our work, we also enable \emph{online} process analysis--streaming process mining~\cite{burattin2022streaming}--as feedback at runtime is crucial for activity executions and decision making in IoT (Objective~2). As we aim to derive abstracted, high-level events from CPS sensors to inform about (business) process activity executions, we will focus on pattern-based sensor data processing following a more light-weight \emph{pipes and filters} style~\cite{khezemi2024systematic} (Objective~3). The IoT systems serving as running examples in this work (cf.~Sections~\ref{sec:setup-manufacturing} and~\ref{sec:setup-healthcare}) indicate that relevant patterns to detect starts and ends of process activities might be referring to events from a rather small number of sensors and their combinations~\cite{kirikkayis2023integrating}, which would make a pattern matching approach based on domain knowledge feasible, especially for runtime detections~\cite{klein2013using}. We will also discuss how more complex sensors (e.g., cameras~\cite{malburg2021object}) and sensor networks can be integrated into our activity detection approach following a pre-processing and abstraction phase to represent more realistic IoT settings~\cite{seiger5165943online}.
 
\subsection{Domain-specific Languages for Activity Detection}

We acknowledge that a large corpus of work related to DSLs~\cite{kosar2016domain,iung_systematic_2020} and their engineering~\cite{krahn2010monticore,kleppe2008software}, also in the context of CPS~\cite{butting2023towards,wood2021triton,pradhan2015chariot} and IoT~\cite{salihbegovic2015design,erazo2022domain}, already exists. The main intent of a DSL is to allow domain experts to specify some kind of data or control-related operations based on the available domain knowledge, not requiring extensive programming skills (Objective~1) or existing data and resources for model training (Objective~3)~\cite{hudak1997domain}. 

DSL-based approaches supporting the general modeling of components, their interactions, and data flow in CPS and IoT systems are presented in~\cite{tichy2020experiences,pradhan2015chariot,salihbegovic2015design}, and with a special focus on monitoring of IoT components in~\cite{erazo2022domain}. More specialized DSLs in the these contexts feature for example the modeling of hazards and risks in CPS~\cite{petzold_pasta_2023}, job scheduling in CPS~\cite{wood2021triton}, or simplified data flow and rule modeling for specific IoT-based condition and location monitoring use cases~\cite{choosumrongdevelopment,phasinam2025real}. These languages feature aspects that we can adapt to model the IoT system and its components (e.g., the grouping of sensors and actuators) in a sensor-agnostic way (Objective~4) and they foster understandability as patterns and rules become explicit (Objective~1). However, none of the DSLs is able to fulfill the functional requirement to perform the necessary event abstraction step and they do not support BPM-related concepts, which is also the case for many approaches that aim at abstracting low-level logs (e.g., containing interaction traces) to derive higher level activities (e.g., in software development processes~\cite{caldeira2016software}). Complex event processing (CEP) has proven to be a suitable, light-weight technology (Objective~3) for abstracting low-level data to higher level events with runtime capabilities (Objective~2), also in the BPM domain according to Soffer et al.~\cite{soffer2019event}. CEP-based platforms usually feature their own SQL-like language to specify the event processing rules and applications. The authors in~\cite{rosa-bilbao_model-driven_2022} argue that CEP languages are still too complex and require technical expertise to be usable by domain experts. Various approaches propose simplifications of these language via abstraction and graphical composition. In~\cite{boubeta-puig_model-driven_2014} Boubeta-Puig et al.~propose a model-driven approach to facilitate the user-friendly design and composition of generic CEP patterns using a visual representation. A visual framework for data flow programming in IoT based on CEP is introduced in~\cite{gokalp2019visual}. 

Even though it is a relevant research problem~\cite{janiesch2020internet}, none of the aforementioned DSLs and approaches share the functional objective of detecting generic business process activities from arbitrary sensors. The approaches closest to this goal can be found in~\cite{negrete_ramirez_dsl-based_2021} proposing a DSL for human activity detections in smart homes and in~\cite{barricelli_visual_2017} with a visual language to detect human activities based on IoT data. While we adapt some patterns and rules from these languages, we acknowledge that there is a need to develop a DSL which is agnostic to a specific application domain, which works with arbitrary sensor data and independent of specific activity types, and which features BPM-related concepts as first class citizens to abstract the data for subsequent process mining analysis. Moreover, targeting domain experts with a rather technical IoT background, we will focus on developing \emph{textual} DSL of efficient sensor pattern specification, rather than a visual notation~\cite{gronninger2014textbased}.

\section{Domain-specific Language: Radiant} \label{sec:dsl}

Addressing the research question~\emph{RQ1}, we present the textual domain-specific language (DSL) \emph{Radiant}, which allows domain experts to specify event abstraction patterns in sensor data to detect activity executions as the main functional objective. Radiant has been developed following the principles and concepts for domain-specific event processing languages laid out in~\cite{bruns2014ds} and in~\cite{karsai2014design}. 

\subsection{Meta-model of Radiant} \label{sec:meta}

\begin{figure}
  \centering
  \includegraphics[width=1.0\linewidth]{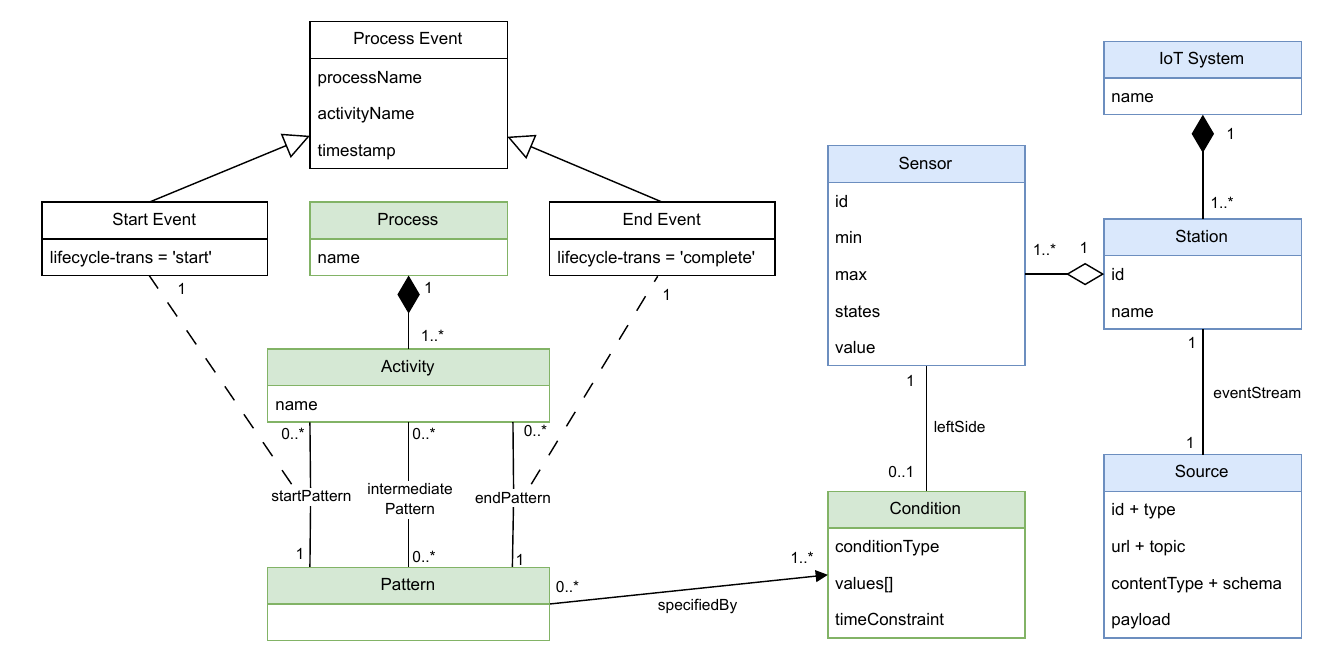}
  \caption{Meta-model of Radiant in UML (green: concepts supported by the DSL; blue: IoT configurations in external YAML file; white: additional BPM concepts)}
  \label{fig:radiant-meta}
\end{figure}

The meta-model of the Radiant language is based on the event-centric meta-model for IoT driven process monitoring presented in~\cite{franceschetti2023event}. As depicted in Figure~\ref{fig:radiant-meta}, a \emph{Process} contains one or more \emph{Activities}~\cite{dumas2013fundamentals}, which we aim to detect. For simplicity reasons, we do not consider nested process structures (i.e., subprocesses) in the current version of the meta-model. An \emph{Activity} is associated with exactly one \emph{startPattern}, exactly one \emph{endPattern}, and an arbitrary number of \emph{intermediatePatterns}. The start and end patterns delimit the occurrence of an activity by a \emph{Start Event} and \emph{End Event} as \emph{Process Events}, which are to be emitted denoting their occurrence (cf.~Figures~\ref{fig:production-process} and~\ref{fig:blood-process}) and used as basis for process mining in subsequent data analyses. A \emph{Pattern} is specified by one or more \emph{Conditions} that have to be fulfilled for it to be detected. A \emph{Condition} refers to the value(s) of one \emph{Sensor}.

Furthermore, the meta-model supports the representation of the IoT system, which is necessary for technical configurations of the runtime system (cf.~Section~\ref{sec:runtime}) and DSL support. An \emph{IoT System} is composed of one or more \emph{Stations}, which aggregate their associated \emph{Sensors}. A \emph{Station} is also associated with a \emph{Source} representing the data stream that contains the station's sensor values as payload. To avoid overloading the domain expert with these technical configuration issues, we decided to separate the configuration of the IoT system from the concepts integrated into concrete syntax of the DSL.

\subsection{Syntax of Radiant} \label{sec:syntax}

Based on the presented meta-model, we now explain the concrete syntax of Radiant in more detail. Note that the following listings presenting the grammar of Radiant are based on the grammar language of Langium\footnote{\url{https://langium.org/}}, which was used as a modern, extensible and flexible framework to implement the DSL~\cite{bork2023language}.

\paragraph{Processes and Activities} The main concepts of Radiant that were adapted from the BPM domain are that of a \emph{Process} and an \emph{Activity}. In Radiant, the process represents the context to which an activity belongs. An arbitrary number of activities (at least one) to be detected can be contained in a process (cf.~Listing~\ref{lst:syntax-process}). As the process concept here only contextualizes an activity, the activity ordering is of no relevance and there are no dependencies among activities. Process discovery and conformance checking are subject to later analysis steps in process mining, which are out of scope of this work.

\begin{lstlisting} [xleftmargin=3.2cm,language=radiant,float,linewidth=0.75\columnwidth,caption={Radiant syntax: Process with Activities; Activity with Patterns}, label={lst:syntax-process}]
'Process' name=ID ':'
  (activities+=Activity)*;

Activity:
  'Activity' name=ID ':'
    startPattern=Start
    (intermediates+=Intermediate)*
    endPattern=End;
\end{lstlisting}

\paragraph{Patterns} An \emph{Activity} is characterized by a number of \emph{Patterns} within the IoT sensor data that have to be detected, most importantly a \emph{startPattern} and an \emph{endPattern}, which are both mandatory (cf.~Listing~\ref{lst:syntax-process}) as they represent the relevant process-level events to be used for process mining (cf.~Section~\ref{sec:preliminaries}). An arbitrary number of \emph{intermediatePattern}s can be specified in between the start and end pattern to make the activity detection more precise and robust. This might be needed to resolve potential ambiguities when more than one activity is characterized by the same patterns~\cite{franceschetti2023characterisation}. If the domain expert is aware of relevant sensor patterns in the data, they should be specified as intermediate pattern to make the detection robust and progress trackable. The patterns defined to be part of an activity are all dependent in the order of their specification from the start pattern, to intermediate patterns, to the end pattern--allowing for stateful processing.

\paragraph{Conditions} Each pattern consists of one or more \emph{Condition}s specifying when the pattern is detected based on the different types (Start, Intermediate, or End). As shown in detail in Listing~\ref{lst:syntax-patterncase}, we support conjunctions and disjunctions of an arbitrary number of conditions. Hereby, each \emph{Case} keyword indicates that the conditions should be considered to be linked by a logical OR. Otherwise, and within one Case statement, the list of conditions specified for a pattern are considered to be linked by a logical AND.

\begin{lstlisting} [xleftmargin=2.6cm, language=radiant, float, linewidth=0.82\columnwidth,caption={Radiant syntax: Different Pattern types; Pattern with Conditions and Cases.}, label={lst:syntax-patterncase}]
Pattern:
  Start | Intermediate | End;

Start:
  'Start:'
    (conditions+=Condition+ | cases+=Case+);

Intermediate:
  'Intermediate:'
     (conditions+=Condition | cases+=Case)+;

End:
  'End:'
    (conditions+=Condition | cases+=Case)+;

Case:
  'Case:'
    (conditions+=Condition)+;
\end{lstlisting}

\paragraph{Sensors} The most important concepts adapted from IoT are \emph{sensor}s indicating the values to be analyzed and their location \emph{in} a specific IoT system or device (e.g., production station)~\cite{bauer2013iot}, which is relevant for locating the specific data stream to be processed in our setup. These concepts are the base of a \emph{condition}, which is displayed in Listing~\ref{lst:syntax-base}. This base is complemented by one mandatory condition type from the ones listed in Table~\ref{tab:conditions}. Radiant currently supports these 10~different types as they cover the detection of all process activities we encountered in our use cases and discussions with domain experts. This list can be extended easily. Conditions that indicate a specific change (types ChangeCondition, ChangingCondition, IncreasingCondition, DecreasingCondition) can optionally be extended with a \emph{time constraint} defining a time-based window within which the change must happen to fire a higher-level event~\cite{jain2008towards,etzion2010event}. The \emph{ChangingCondition} refers to any change in the specific sensor value, which can be made more coarse-grained via discretization (cf.~Section~\ref{sec:configs}).

\begin{lstlisting} [xleftmargin=2.2cm, language=radiant,float,linewidth=0.85\columnwidth,caption={Radiant syntax: Condition Base with optional time constraint.}, label={lst:syntax-base}]
Condition:
  'In' station=ID 'sensor' sensor=ID ConditionType (time_constraint+=TimeConstraint)*

TimeConstraint:
  'within' amount=INT time_unit=ID;
  
\end{lstlisting}

\begin{table}[]
\caption{Condition Types supported by Radiant}
\centering
\small
\begin{tabular}{|l|l|}
\hline
 \textbf{Condition Type} & \textbf{Syntax} \\ \hline
ChangeCondition & changes\_from \textit{Value} to \textit{Value} \\
 &\qquad (TimeConstraint)*; \\ \hline
 ChangingCondition & is\_changing (TimeConstraint)*; \\ \hline
RangeCondition & in\_range \textit{Value} to \textit{Value}; \\ \hline
IsEqualCondition & is\_equal \textit{Value}; \\ \hline
IsLowerCondition & is\_lower \textit{Value}; \\ \hline
IsLowerOrEqualCondition & is\_lower\_or\_equal \textit{Value}; \\ \hline
IsHigherCondition & is\_higher \textit{Value}; \\ \hline
IsHigherOrEqualCondition & is\_higher\_or\_equal \textit{Value}; \\ \hline
IncreasingCondition & is\_increasing (TimeConstraint)*; \\ \hline
DecreasingCondition & is\_decreasing (TimeConstraint)*; \\ \hline
\end{tabular}
\label{tab:conditions}
\end{table}

\subsection{IoT Configurations} \label{sec:configs}

\paragraph{Sensors} When working with rather low-level sensor data from IoT (e.g.,~as shown in Listings~\ref{lst:sorting} and~\ref{lst:donor}), the requirement of sensor discretization emerged to make the specification of the conditions more user friendly and more robust against variations in continuous data (e.g., by considering ranges in the sensor discretization, as opposed to exact values that need to be met)~\cite{mangler2024internet}. Using descretization, change-related conditions from Table~\ref{tab:conditions} can also be made more coarse-grained to account for sensors that show a high frequency of insignificant changes (e.g., environmental sensors with high precision); thus, to indicate changes in the patterns of the discretized sensor values that are more relevant for activity detection.

Moreover, available sensors, their data types, value ranges, and associations with IoT devices have to be pre-configured to make the DSL more usable and support the domain expert with auto-suggestions, validations (e.g., sensor values being out of range) and code completions--increasing productivity~\cite{kosar2008preliminary}. While these aspects might also be part of the DSL, we decided to treat them as a separate, more technical concern and externalize it into an external YAML-based configuration file following the structure and attributes of entities presented in the meta-model (cf.~Section~\ref{sec:meta}). Listing~\ref{lst:config} shows parts of this sensor configuration for the sorting machine station. Here we see the definition of a template for the motor speeds (lines~1--8) with state abstractions and min/max values that can be reused in the specification of the actual motor sensors (line~24) as all motors in the smart factory have this same behavior. The listing also contains an exemplary discretization of the Integer values emitted from the color sensor (lines~18--21).

\begin{lstlisting} [xleftmargin=2.4cm, float, linewidth=0.82\columnwidth,caption={Sensor and source stream configurations in a YAML file.}, label={lst:config}]
presets:
  - id: motor_preset
    min_value: -512
    max_value: 512
    states:
      low: -512
      off: 0
      high: 512

stations:
  - id: SM_1
    name: Sorting Machine
    source: SM_1Stream
    sensors:
      - id: i1_light_barrier
        type: switch
      - id: i2_color_sensor
        discretization:
          lower: [1725, "red"]
          intermediate: [1725, 1790, "blue"]
          upper: [1790, "white"]
      - id: m1_speed
        type: int
        preset: motor_preset
      - id: o5_valve
        type: int
        states:
          open: 75
          closed: 0
          
sources:
  - id: SM_1Stream
    type: mqtt
    url: ${MQTT_URL}
    client_id: mqtt.SM_1.Sort
    topic: FTFactory/SM_1
    content_type: json
    schema:
      ts: string
      i1_light_barrier: int
      i2_color_sensor: int
      m1_speed: int
      o5_valve: int

\end{lstlisting}

\paragraph{Source and Sink Streams}

The \emph{sources} and \emph{sinks} of the sensor data streams need to be configured, as well. Our goal is to detect activity executions at runtime from streams of IoT sensor data. We assume that typical messaging systems and brokers (e.g.,~MQTT, RabbitMQ, Apache Kafka, etc.) are available as sources to emit these sensor events and we can consume them for analysis in a publish-subscribe manner~\cite{corral2020stream}. The configurations to connect to these systems and map the event data to the internal event processing data models are also part of the external configuration file. For the \emph{sinks} of the event processing, we suggest to emit the process-level start and end events from the activity detection in a standardized format for subsequent process analysis (e.g.,~in XES format~\cite{gunther2014xes}) on one stream per process instance~\cite{burattin2020mqtt}. Listing~\ref{lst:config} also presents an exemplary source stream configuration (lines~31--43) to access the event streams from the sorting machine station of the smart factory via MQTT, which is referred to via the ID in the \emph{source} parameter of the sensor configurations (cf.~listing~\ref{lst:config}, line~13). In addition, the listing contains the schema definition of the JSON-based messages received on this event stream (lines~38--43), which is used to map the payload to the internal event model. 

\subsection{Radiant Examples} \label{sec:radiant-examples}

\begin{lstlisting} [xleftmargin=2cm, language=radiant, float, linewidth=0.85\columnwidth,caption={Radiant example for activities of the production process.}, label={lst:syntax-example2}]
Process Production:
  Activity Mill_workpiece:
    Start:
      In MM_1 sensor i1_pos_switch is_equal 1;
      In MM_1 sensor o8_compressor
        changes_from off to on;
    End:
      In MM_1 sensor m1_speed changes_from 512 to 0;

  Activity Sort product:
    Start:
      In SM_1 sensor m1_speed 
        changes_from 0 to -512;
      In SM_1 sensor i1_light_barrier is_equal 1;
    Intermediate:
      In SM_1 sensor i2_color_sensor is_changing;
    End:
      Case:
        In SM_1 sensor o5_valve 
          changes_from open to closed;
      Case:
        In SM_1 sensor o6_valve 
          changes_from open to closed;
\end{lstlisting}

\paragraph{Production Process}
In Listing~\ref{lst:syntax-example2} we present an example for activities of the production process (cf.~Section~\ref{sec:setup-manufacturing}) created with Radiant. The start of the \emph{Mill workpiece} activity in the milling machine (MM\_1) is indicated by the state of a position switch (line~4) and the change of a compressor from state \emph{off} to \emph{on} (lines~5--6) as defined in the sensor configuration file. The end of the milling activity (lines~7--8) is specified as the speed of a specific motor decreasing to zero (i.e., coming to a halt). The patterns to detect the \emph{Sort product} activity executed by the sorting machine (SM\_1) are shown in Listing~\ref{lst:syntax-example2}, lines~10--23. Note that here we define one additional intermediate pattern to track the progress of the activity execution by observing a change in the color sensor readings (lines~15--16). The activity's end is marked via a logical OR specifying that one of the available valves (cf.~Figure~\ref{fig:factory-iot}) changes its state from \emph{open} to \emph{closed}--depending on the previous color reading--and thus performed the actual sorting (lines~17--23).

\paragraph{Blood Donation Process} 
In Listing~\ref{lst:syntax-example1} we present a Radiant example for parts of the blood donation process (cf.~Section~\ref{sec:setup-healthcare}). The start of the \emph{Sanitize hands} activity (line~2) is detected when the load cell sensor in the central hygiene station (CHS) changes its state from low to high (lines~4--5). This pattern requires a discretization of the sensor values emitted from the load cell (in gramms) to more abstract states. In our setup, we assume that the bottle with the hand sanitizer is placed on the load cell and it has an initial weight which will decrease with every usage. We denote this state with slowly decreasing weights as \emph{low}. When the healthcare worker applies pressure to dispense liquid from the sanitizer, we can observe a significant brief peak in the load cell's values ($>$ 1000~g) in a value range we denote as \emph{high}. Similarly, we detect the end of this activity based on a change from high to low (lines~6--8). Note that for both conditions, we specify a time window constraint of 30 seconds as we assume that the healthcare worker might use the dispenser several times to perform one instance of the \emph{Sanitize hands} activity.

\begin{lstlisting} [xleftmargin=2cm, language=radiant, float, linewidth=0.85\columnwidth,caption={Radiant example for an activity of the blood donation process.}, label={lst:syntax-example1}]
Process Blood_donation:
  Activity Sanitize_hands:
    Start:
      In CHS sensor load_cell 
        changes_from low to high within 30 seconds;    
    End:
      In CHS sensor load_cell 
        changes_from high to low within 30 seconds;
\end{lstlisting}

\subsection{Code Generation}

The core requirement of the target language and execution platform should be its ability to process sensor data from one or more event streams and detect specific patterns in the current events as well as in subsequent events. Complex event processing (CEP) is the most suitable technology here as it supports exactly these features~\cite{etzion2010event}. As a concrete CEP platform, we decided for \emph{Siddhi} which is part of the WSO2 stream processing platform~\cite{suhothayan2011siddhi}. Siddhi is a light-weight Java-based service optimized for high-velocity and high-throughput event processing~\cite{perera2014solving}. It features its own event processing language called \emph{StreamingSQL}~\cite{jain2008towards}, which is the main target (host) language of our implemented code generator to translate the Radiant artifacts to. 
StreamingSQL supports all concepts required to detect patterns in sensor event streams w.r.t.~the corresponding activity executions.

As pointed out in~\cite{rosa-bilbao_model-driven_2022}, CEP languages are still too complex and require technical expertise to effectively write event processing applications. Here, following a model-driven engineering approach~\cite{rosa-bilbao_model-driven_2022}, Radiant introduces necessary abstractions to support domain experts with writing applications for the specific use case of activity detection. The Siddhi CEP platform includes a runtime system where the generated \emph{Siddhi apps} can be directly deployed to and executed via a REST API, thus allowing seamless development workflows from composition with Radiant to execution and activity detection at runtime from live IoT sensor data. If necessary, domain experts can also inspect and modify the generated Siddhi apps before deployment in a web-based editor integrated with the Siddhi CEP platform.

\begin{lstlisting} [linewidth=0.97\columnwidth,float, caption={Example of a Siddhi CEP app (in StreamingSQL) generated from a Radiant application for detecting one type of process activity.}, label={lst:siddhi-fragments}]
@App:name('Production-Sort product')

@source(type = 'mqtt', url = '${MQTT_URL}', topic = 'FTFactory/SM_1', @map(type = 'json'))
define stream SM_1Stream(ts string, i1_light_barrier int, i2_color_sensor int, m1_speed int, o5_valve int, o6_valve int);

@sink(type = 'log')
define Stream DetectedActivities(event string, activity string, ts_start string, ts_end string);

@info(name='StartPattern')
from every e1 = SM_1Stream, e2 = SM_1Stream[(e1.m1_speed==0 and e2.m1_speed==-512) and (e2.i1_light_barrier==1)] 
select 'Start' as event, 'Sort product' as activity, e2.ts as ts
insert into DetectedPatterns;

/* ... */

@info(name="EndPattern")
from every e1 = SM_1Stream, e2 = SM_1Stream[((e1.sm1_o5_valve==75 and e2.sm1_o5_valve==0)) or ((e1.sm1_o6_valve==75 and e2.sm1_o6_valve==0))] 
select "EndPattern" as event, "Sort product" as activity, e1.ts as ts
insert into DetectedPatterns;

@info(name="Detect-Activity")
from every e1 = DetectedPatterns[event == "StartPattern"] -> not DetectedPatterns[event == "StartPattern"] and e2 = DetectedPatterns[event == "IntermediatePattern"] -> not DetectedPatterns[event == "StartPattern"] and e3 = DetectedPatterns[event == "EndPattern"]
select "Sort product" as activity, e1.ts as ts_start, e3.ts as ts_end
insert into DetectedActivities;

\end{lstlisting}

\paragraph*{Example} Listing~\ref{lst:siddhi-fragments} presents exemplary fragments of a Siddhi app generated from the Radiant example for the \emph{Sort product} activity in the production process (cf.~Listing~\ref{lst:syntax-example2}). We generate \emph{one app per activity} that is defined in a Radiant application with the process name providing the execution context of an activity. We decided for a one-to-one mapping of activity to Siddhi app as this leads to cohesive, standalone apps that do not have other dependencies. This granularity is driven by the problem domain--the detection of activities--and allows us to easily extend the set of supported activities. Note that this corresponds to the overarching goal of this work related to the event abstraction problem that precedes process mining~\cite{diba2020extraction}. The analysis of the interrelations of detected process events and activities is subject to process mining analysis~\cite{van2012process}, which is out of scope of this work. For each station with its sensors used in a Radiant application, we additionally retrieve the IoT configuration parameters from the YAML file (cf.~Section~\ref{sec:configs}) during code generation to add these configurations to the beginning of the generated Siddhi apps.

We can find the definition of the source stream that connects to events emitted from the IoT system via MQTT in lines~3--4. Note that with our setup for the smart factory (cf.~Section~\ref{sec:setup-manufacturing}), we assume that each production station emits messages on one dedicated source stream. Each message contains the status of all sensors for one station at one point in time as attributes. We translate the provided patterns into the corresponding patterns of StreamingSQL referring to the source event stream(s) specified in the Radiant application. In case an activity comprises multiple stations (e.g., in the healthcare setup), thus multiple source event streams, the events from these two different streams first have to be joined and added to a new event stream to access all relevant event attributes~\cite{jain2008towards}. An exemplary sink stream where detected activities events are emitted to is defined in lines~6--7. For simplicity, we only \emph{log} the event occurrence, more sophisticated event sinks (e.g.,~using MQTT-XES~\cite{burattin2020mqtt}) can be added via the configuration file.

Lines~9--19 in Listing~\ref{lst:siddhi-fragments} contain the translated StreamingSQL queries to detect the start pattern and end pattern in the sensor data (skipping intermediate patterns) for the \emph{Sort product} activity specified in Listing~\ref{lst:syntax-example2}. For the start and end patterns, the changes in the attributes of two subsequent events $e1$ and $e2$ on the same event stream are analyzed for the specified conditions. We support the translation of all conditions shown in Table~\ref{tab:conditions} into the concrete StreamingSQL syntax. If all conditions specified in one query are met, a new high-level event is inserted into a new event stream (here for detected patterns). Note that in line~22 we can find a more complex, automatically generated query which tracks the occurrence of all specified patterns in the sequence from start to end. With this we track the progress of the pattern detection and also ensure that the event indicating the \emph{end} of an activity execution is only emitted when corresponding start and all the previous intermediate patterns have occurred in their specified sequence.  The event selected in line~23 contains all the information that we ultimately aim to derive. 

\subsection{Architecture of Runtime System} \label{sec:runtime}

\begin{figure}
  \centering
  \includegraphics[width=0.7\linewidth]{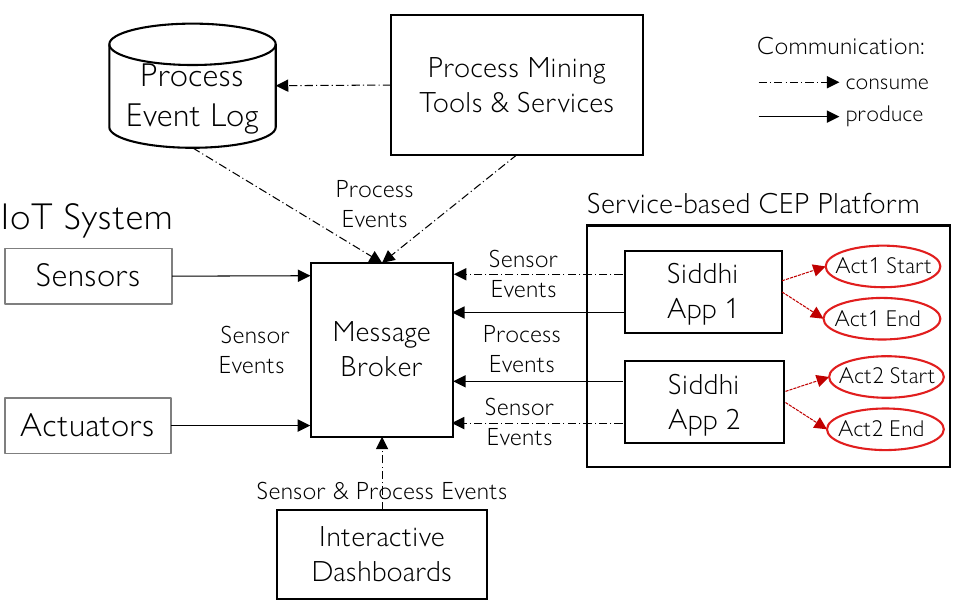}
  \caption{Software architecture: CEP apps consuming sensor events and producing process events to be consumed by other software components.}
  \label{fig:architecture}
\end{figure}

Addressing the research question~\emph{RQ2}, we present a software architecture that enables the event abstraction from IoT data at runtime. Figure~\ref{fig:architecture} shows the overall architecture of the CEP-based process activity detection system at runtime. The Radiant applications are translated into \emph{Siddhi apps}, which are deployed and executed on the service-based CEP platform \emph{Siddhi} serving as light-weight runtime environment. In our proposed architecture, one Siddhi app is capable of detecting one type of activity, emitting process-level start and end events when an activity execution has been detected. This way, we can flexibly add new Siddhi apps for new types of activities, and easily activate/deactivate these apps for specific types of activities at runtime using the REST API of the Siddhi \emph{Runner} service. Furthermore, we can leverage distributed network architectures by deploying specific activity detection apps on different light-weight CEP platforms running closer to the edge (e.g., as part of one a station of the IoT system) for local sensor pre-processing and only emitting abstracted, higher level events~\cite{seiger5165943online,systa2023liquidai}. As the main goal of our work is to provide an abstraction on top of IoT data in the form of activity executions, we expect that the number of process activities to be detected simultaneously does not need to scale significantly and that there is no need for a more fine-grained partitioning.

\begin{figure}[t!]
  \centering
  \frame{\includegraphics[width=1.00\linewidth]{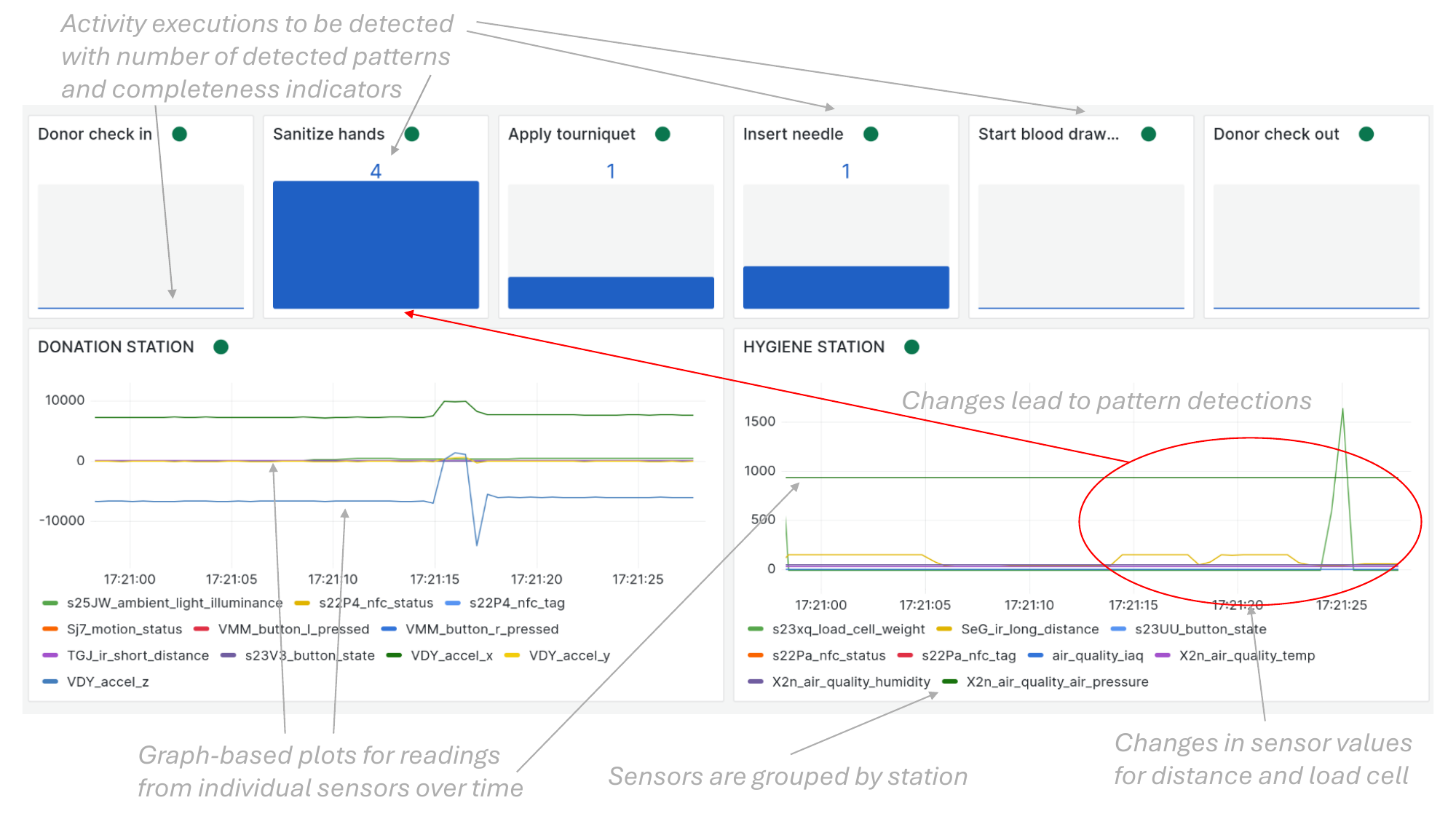}}
  \caption{Visualization of activity detections and sensors in interactive dashboard, with annotations for activity detections (top) and corresponding IoT data streams (bottom).}
  \label{fig:dashboard}
\end{figure}

The IoT devices produce sensor events to be sent on one or more topics to the message broker. The Siddhi apps are subscribed to these topics to consume and process the incoming events independently from each other according to the stream configurations generated from the external configuration file. The process-level events produced and emitted from the Siddhi apps can then be either persisted in a process event log (e.g., via a subscribed persistence or logging service) for offline process mining and/or consumed and processed in streaming process mining cases by other process mining tools and services~\cite{burattin2022streaming}. In case of running the event processing on an edge device with sensors directly attached, the message broker might be replaced by a more light-weight system/way to ingest the sensor data locally. Furthermore, we setup interactive dashboards that are subscribed to process and sensor events for visualization and exploration. Figure~\ref{fig:dashboard} shows an annotated screenshot of a Grafana-based dashboard which visualizes the number of detected subsequent low-level patterns for a process activity as a bar chart (top) and the associated low-level IoT sensor data as a time series (bottom). In the example, changes in the distance sensors followed by changes in the weight values of the load cell sensor triggered multiple patterns that are part of the \emph{Sanitize hands} activity.

As the architecture is event-driven and service-based~\cite{seiger2023data}, other more complex IoT sensors that may need additional pre-processing steps can be integrated, which makes the approach also applicable to more realistic IoT settings outside lab environments. Here, a basic requirement is that the corresponding software components (services) perform a processing of the complex raw data to more abstract event and (virtual) sensor data~\cite{martin2021virtual} that can be considered as part of the pattern and condition specification in Radiant (e.g., the abstract event \emph{machine dropped workpiece} determined by a camera-based object detection service~\cite{malburg2021object}). These abstracted sensor events and corresponding event streams just need to be added to the external configuration file (cf.~Section~\ref{sec:configs}). 

\subsection{Implementation}

Radiant has been implemented using the \emph{Langium} framework which uses the Language Server Protocol~\cite{bork2023language}. We decided for Langium to have a modern, flexible platform for language development that decouples the front-end from the language implementation~\cite{bunder2019decoupling}. We have exemplary integrations with Visual Studio Code and NeoVim supporting the user via syntax highlighting, auto-suggestions, code completion and verifications based on the configuration files. The implementation of Radiant can be found on GitHub\footnote{\url{https://github.com/ics-unisg/radiant-iot-activity-dsl}} and as Visual Studio Code extension\footnote{\url{https://marketplace.visualstudio.com/items?itemName=mahgoh.radiant}}. We have implemented a code generator to translate the Radiant applications into Siddhi apps~\cite{suhothayan2011siddhi}. As Langium  has a strong focus on extensibility, new generators to translate into code for other CEP platforms (e.g., Apache Storm) can be added.

\section{Evaluation \& Discussion} \label{sec:evaluation}

The evaluation of the Radiant applications and the CEP-based activity detection using these applications is based on the example processes from smart manufacturing (cf.~Sect.~\ref{sec:setup-manufacturing}) and smart healthcare (cf.~Sect.~\ref{sec:setup-healthcare}). The goal of this evaluation is to demonstrate the activity detections based on CEP apps generated from Radiant applications that were composed by domain experts for the two IoT scenarios. Through this, we aim to demonstrate the feasibility and validity of Radiant and the corresponding architecture to be used as means for IoT-based process event abstraction at runtime.

\subsection{Experimental Setup}
To demonstrate the validity and feasibility of the DSL and its runtime capabilities, we asked domain experts familiar with the laboratory setups to write the Radiant applications for all relevant activities in the scenario processes (cf.~Section~\ref{sec:preliminaries}). \textcolor{black}{Both the expert from the medical domain and the expert from the manufacturing domain have been involved in the discussions of important concepts and patterns to be integrated into Radiant in earlier development stages (cf.~Section~\ref{sec:dsr}). They got a brief introduction to the final language concepts and were then tasked with specifying the activity-related patterns. Both experts did not have prior skills in programming or other forms of querying/pattern languages, but they did have a basic understanding of computer science concepts, which was helpful, e.g., with correctly specifying conjunctions and disjunctions of patterns. The YAML-based configuration files for the two setups were provided by the responsible IoT engineers who are familiar with the laboratory setups and their existing sensors.} The Radiant applications were translated into Siddhi apps and deployed to the runtime system. We executed several instances of the processes, recorded the sensor data, and logged the process-level events of the activity detections. The processes in the smart factory were orchestrated by a BPM system~\cite{seiger2022integrating}, which generates an event log of activity executions that is used as ground truth to compare the Radiant-based activity detection with. For the executions in the smart healthcare setup, we relied on a manual monitoring and logging of the activity executions to create the ground truth. 
The dataset with all Radiant applications, configuration files, generated Siddhi apps, IoT sensor data, and ground truth logs can be found in~\cite{seiger_2025_15773369}. 

\subsection{Performance}
The Siddhi CEP engine has proven its capabilities of filtering and processing more than one million events per second in memory and more than 300.000 events per second via network~\cite{perera2014solving}. These benchmarks make the engine a suitable technology for larger scale, distributed IoT deployments and event abstraction at runtime.

During our experiments capturing, replaying and processing the IoT data streams, we did not experience any performance-related issues on state-of-the-art laptop computers. We were able to run 14~different CEP applications for activity detection in parallel in our laboratory setup, processing 1200 IoT sensor readings per minute (limited by the sampling frequency of the sensors)~\cite{seiger5165943online}. This confirms that our current choices of granularity (i.e., one Siddhi app per activity) and one instance per app on the CEP platform are feasible in the given IoT setups. 

An exemplary runtime resource consumption on our experimentation laptop (Intel Core i7-8550U@1.8GHZ, 4 Cores, 16 GB RAM, Windows 11) running the Java-based Siddhi Runner (v5.1.2\footnote{\url{https://siddhi.io/en/v5.1/download/}}) application in a 64-bit JVM (Java~1.8) with 14~active Siddhi apps (generated from Radiant) included CPU loads between 0.2\% and 1\% and main memory consumption of 9.4\,MB per Siddhi app with active connections to 31~sensors via MQTT; each sensor sampling its data at 2\,Hz. The memory consumption of the Siddhi Runner including the JVM it runs in, without any active CEP apps was 224\,MB on average. Latency measurements resulted in average event processing times of less than 50\,ms, which include sampling a sensor, sending its data to an MQTT broker on a different computer via Gigabit-Ethernet, consuming it from the MQTT broker via the corresponding Siddhi app, processing the sensor data, and logging the occurrence of a process-level event. Large-scale performance benchmarks with high-frequency IoT sensors and devices remain subject to future work.

\subsection{Detection Metrics}
The comparison of the activity detection logs with the ground truth was performed using the \emph{AquDem} tool presented in~\cite{kurz2024activity}. We use comparison metrics from several categories that are supported by AquDem as shown in Tables~\ref{tab:overall-fact} and~\ref{tab:overall-hc}. The first two, devised by Ward et al.~\cite{wardmetr2011}, are based on the classification of detections: 1)~\emph{Two Set} metrics are frame-based, which means they compare individual frames from the ground truth with those from the detection and classify them into different categories (e.g., true positives, true negatives, deletions, fragmentations, mergings, etc); 2)~\emph{Event Analysis} metrics are based on the Two Set metrics, using them to categorize entire activity detections as correct (i.e.,~a detected event corresponds to the ground truth event), deleted (i.e.,~an event from the ground truth was not detected), fragmented (i.e.,~one event in the ground truth is recognized by several detected events), merged (i.e.,~several ground truth events correspond to one detected event), etc~\cite{wardmetr2011}. The classifications of frames and activities provided by these metric groups can then be used to provide a range of standard metrics, e.g., precision ($Precision = \frac{TP}{TP + FP}$), recall ($Recall = \frac{TP}{TP + FN}$), the F1 score ($F1 = \frac{2 * Precision * Recall}{Precision+ Recall}$)~\cite{Taha2015} and balanced accuracy ($BA = \frac{1}{2}(\frac{TP}{TP + FN}+ \frac{TN}{TN + FP})$)~\cite{mosley2013balanced}. Notably, for entire activity classifications no concept of true negatives exists and balanced accuracy is therefore not reported for the Event Analysis metrics. Furthermore, \emph{cross-correlation} measures the similarity between the detected and ground truth time series by determining the time shift at which the corresponding frames are most alike and then quantifying that similarity~\cite{lyon2010discrete}. Finally, the normalized \emph{Damerau-Levenshtein distance} measures the dissimilarity between the ground truth and detection activity sequence in number of edits that would be necessary to make the sequences equal~\cite{damerau_technique_1964}, normalized by the length of the longer sequence.

\subsection{Experiments and Results: Smart Manufacturing}

\begin{table}[h!]
\centering
\small
\caption{Overall activity detection metrics for the smart manufacturing scenario, micro-averaged over all activities.}
\begin{tabular}{l|llll}
\textit{Metric Category}                                       & \multicolumn{4}{l}{\textit{Metrics}}                                                                                                                                                \\ \hline \hline
\multicolumn{1}{|l|}{\multirow{2}{*}{Two Set}}         & \multicolumn{1}{l|}{\textbf{Precision}}        & \multicolumn{1}{l|}{\textbf{Recall}} & \multicolumn{1}{l|}{\textbf{F1}} & \multicolumn{1}{l|}{\textbf{Bal Acc}} \\ \cline{2-5}
\multicolumn{1}{|l|}{}                                & \multicolumn{1}{l|}{0.3482}                      & \multicolumn{1}{l|}{0.5956}            & \multicolumn{1}{l|}{0.4395}        & \multicolumn{1}{l|}{0.7513}                       \\ \hline \hline
\multicolumn{1}{|l|}{\multirow{2}{*}{Event Analysis}} & \multicolumn{1}{l|}{\textbf{Precision}}        & \multicolumn{1}{l|}{\textbf{Recall}} & \multicolumn{2}{l|}{\textbf{F1}}                                                   \\ \cline{2-5} 
\multicolumn{1}{|l|}{}                                & \multicolumn{1}{l|}{0.6276}                      & \multicolumn{1}{l|}{0.6675}            & \multicolumn{2}{l|}{0.6469}                                                          \\ \hline \hline
\multicolumn{1}{|l|}{\multirow{2}{*}{Other}}          & \multicolumn{1}{l|}{\textbf{Damerau-Lev-Norm}} & \multicolumn{3}{l|}{\textbf{Cross-Correlation}}                                                                           \\ \cline{2-5} 
\multicolumn{1}{|l|}{}                                & \multicolumn{1}{l|}{0.4934}                      & \multicolumn{3}{l|}{0.7438}                                                                                                 \\ \hline
\end{tabular}
\label{tab:overall-fact}
\end{table}

\begin{figure}[h!]
  \centering
  \includegraphics[width=0.85\linewidth]{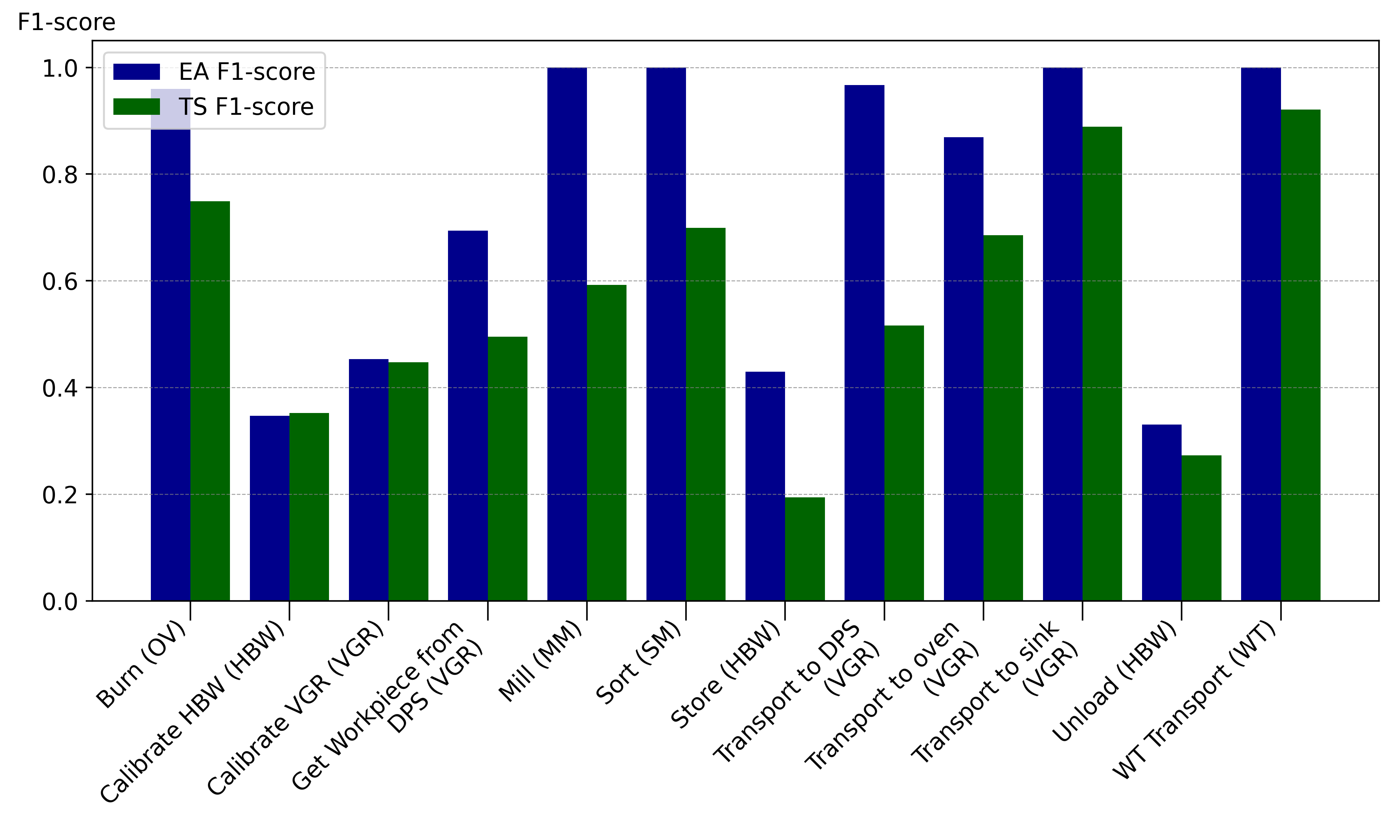}
  \caption{Event Analysis (EA) F1 and Two Set (TS) F1 scores per activity in the smart manufacturing scenario.}
  \label{fig:fact-ea-ts-f1-act}
\end{figure}

We replayed the execution of five recorded IoT sensor logs, spanning in total 181.4 minutes with 424 process activity instances executed. These logs contain the timestamped readings from 77~distinct sensors and actuators, each sampled at 0.5\,Hz. Activities in the smart factory include the transportation of workpieces, storage and unloading, sorting, simulated burning and milling (cf.~Figure~\ref{fig:fact-ea-ts-f1-act}). Their durations were in the range of 5~seconds (\textit{Calibrate VGR}) to 49~seconds (\textit{Unload from High-bay Warehouse}). Relevant evaluation metrics of the detection performance can be found in Table~\ref{tab:overall-fact}. A breakdown of F1 scores for the different activities is shown in Figure~\ref{fig:fact-ea-ts-f1-act}. Note that the F1 scores in Table~\ref{tab:overall-fact}, showing the overall metrics, are lower than the average over the activity-specific metrics in Figure~\ref{fig:fact-ea-ts-f1-act}, since we use \emph{micro-averaging} over the activities and some of the lower performing metrics occur more frequently. Micro-averaging over the activities means that we first sum up the frame or event classifications over all activities and then calculate metrics and rates on these sums~\cite{sokolova_systematic_2009}. If we consider the \emph{macro-average} over the activities, which does not take the number of events or frames for each activity into account, the F1 score for the Two Set metrics is $\sim 0.57$, and for the Event Analysis metrics it is $\sim 0.75$.

\subsection{Experiments and Results: Smart Healthcare}

\begin{figure}[h!]
  \centering
  \includegraphics[width=0.85\linewidth]{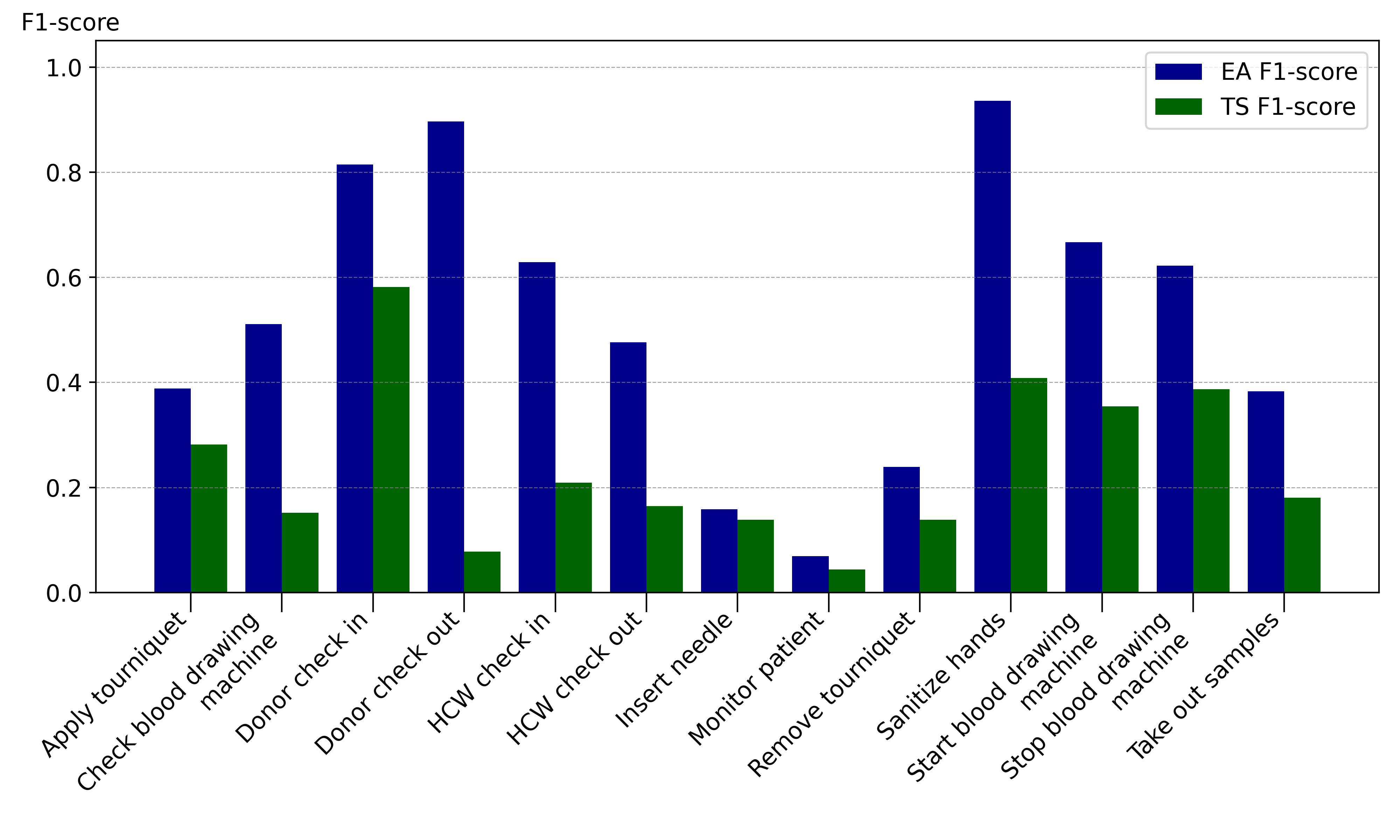}
  \caption{Event Analysis (EA) F1 and Two Set (TS) F1 scores per activity in the smart healthcare scenario.}
  \label{fig:hc-ea-ts-f1-act}
\end{figure}

For the evaluation of the healthcare scenario, we recorded and replayed the execution of 16 IoT sensor logs, each containing one process execution. These processes have been executed by students of the medical master program at our university. In total, the logs span 65.31 minutes and include 240 relevant activity executions. The IoT logs contain 31~distinct, timestamped sensor values, each sampled at 2\,Hz. Activities in the smart healthcare setup include the check-ins of patient and healthcare worker, hand hygiene, applying/remove the tourniquet, inserting/removing the needle and operating the blood drawing machine (cf.~Figure~\ref{fig:hc-ea-ts-f1-act}). Their durations were in the range of 3~seconds (Donor check-in) to 20~seconds (Sanitize hygiene). The detection metrics can be seen in Table~\ref{tab:overall-hc}, and F1 scores per activity in Figure~\ref{fig:hc-ea-ts-f1-act}. Note again the effect of micro-averaging when comparing the values between Table~\ref{tab:overall-hc} and Figure~\ref{fig:hc-ea-ts-f1-act}. The macro-average F1 score over the activities in the healthcare scenario is $\sim 0.24$ for the Two Set metrics and $\sim 0.52$ for the Event Analysis metrics.

\begin{table}[b!]
\centering
\small
\caption{Overall activity detection metrics for the smart healthcare scenario, micro-averaged over all activities.}
\begin{tabular}{l|llll}
\textit{Metric Category}                                       & \multicolumn{4}{l}{\textit{Metrics}}                                                                                                                                                \\ \hline \hline
\multicolumn{1}{|l|}{\multirow{2}{*}{Two Set}}         & \multicolumn{1}{l|}{\textbf{Precision}}        & \multicolumn{1}{l|}{\textbf{Recall}} & \multicolumn{1}{l|}{\textbf{F1}} & \multicolumn{1}{l|}{\textbf{Bal Acc}} \\ \cline{2-5}
\multicolumn{1}{|l|}{}                                & \multicolumn{1}{l|}{0.1358}                      & \multicolumn{1}{l|}{0.3387}            & \multicolumn{1}{l|}{0.1939}        & \multicolumn{1}{l|}{0.6143}                       \\ \hline \hline
\multicolumn{1}{|l|}{\multirow{2}{*}{Event Analysis}} & \multicolumn{1}{l|}{\textbf{Precision}}        & \multicolumn{1}{l|}{\textbf{Recall}} & \multicolumn{2}{l|}{\textbf{F1}}                                                   \\ \cline{2-5} 
\multicolumn{1}{|l|}{}                                & \multicolumn{1}{l|}{0.2643}                      & \multicolumn{1}{l|}{0.7578}            & \multicolumn{2}{l|}{0.3919}                                                          \\ \hline \hline
\multicolumn{1}{|l|}{\multirow{2}{*}{Other}}          & \multicolumn{1}{l|}{\textbf{Damerau-Lev-Norm}} & \multicolumn{3}{l|}{\textbf{Cross-Correlation}}                                                                           \\ \cline{2-5} 
\multicolumn{1}{|l|}{}                                & \multicolumn{1}{l|}{0.725}                      & \multicolumn{3}{l|}{0.7109}                                                                                                 \\ \hline
\end{tabular}
\label{tab:overall-hc}
\end{table}

\subsection{Discussion}

\subsubsection{General Findings}

The results of the evaluation show the validity of the proof-of-concept implementation of Radiant to be used for composing event abstraction applications, translated to corresponding CEP applications, that are capable of detecting process activity executions from IoT data. CEP proves to be a suitable technology allowing effective and efficient pattern detection in IoT data streams. However, the variations of the F1 score demonstrate an important issue we observed when working with process executions and sensors in the physical world: variations in activity executions and the associated IoT data, which show the limitations of pattern-based approaches. The knowledge of domain experts using the DSL hereby plays an essential role to influence and improve the quality of detections by anticipating these potential variations in the sensor data that have to be encoded in the Radiant patterns. For some types of activities, the results show a drop in the detection quality, which is usually due to sensor data not being sufficiently available to distinguish similar activities from each other or to detect them at all. Sections~\ref{sec:disc-sm} and~\ref{sec:disc-sh} discuss these aspects for the two investigated scenarios in more detail. In general, we face a trade-off between increasing the privacy-invasiveness of the monitoring setup using more capable sensors (e.g.,~cameras) to increase the quality of activity detections, but then also requiring more complex pre-processing steps (e.g.,~using computer vision) for abstracting the sensor data to be used in the pattern specifications and sacrificing runtime capabilities. 

In a preliminary informal user study we asked the domain experts who created the Radiant applications for the experiments about their experience. Anecdotal evidence from the domain experts highlights the integrated support through the IDE for efficient ``programming'' of the activity detection patterns. The pattern specification was seen as accessible, also by non-programmers \textcolor{black}{such as the involved domain experts themselves}, and the sensor patterns with the link to process activities were perceived as understandable. Note that a more comprehensive user study with domain experts is required to evaluate Radiant's suitability and usability for specifying process-level events based on IoT patterns. This is subject to future work.

The seamless workflow from writing a Radiant application to the live system detecting activities from IoT sensor data allows us to conduct efficient evaluations of Radiant applications. Given recorded IoT data and a ground truth log, we are able to quickly inform the domain expert about the detection qualities of the composed applications, potential ambiguities and the need for improvements, e.g., by integrating additional patterns and installing additional sensors to increase the monitoring quality, if possible, or to constrain the variability associated with the execution of an activity.
As shown with the experiments, the availability of a pre-existing, recorded IoT sensor log and associated ground truth of activity executions can also help the domain expert in a reverse engineering approach to identify relevant patterns in the sensor data that should be part of the Radiant applications. Based on these preliminary insights, we can hypothesize that Radiant, a DSL serving as an abstraction on top of event processing languages~\cite{boubeta-puig_model-driven_2014}, makes event abstraction using IoT data more accessible to IoT domain experts than other pattern or ML-based approaches, because it integrates important concepts from IoT and BPM as first class citizens (cf.~RQ1 and Objective~1).

Compared to ML-based models and inference, the proposed pattern-based approach is light-weight as it does not require an existing corpus of data and extensive model training (Objective~2). Nevertheless, ML-based sensor processing components can be integrated into the Radiant applications assuming that they emit pre-processed, abstracted sensor events to the message broker according to the proposed runtime architecture (cf.~Section~\ref{sec:runtime}). The applications written by the domain experts and used for the experiments comprise 107~lines of Radiant code for detecting 12~different types of activities in the smart factory processes, and~224 lines of Radiant code for detecting 13~different types of activities the smart healthcare processes. \textcolor{black}{The Radiant application for the smart factory generated 12~Siddhi apps, one for each activity, which are each between 35 to 45 lines of code in size, totaling to 455 lines of Siddhi code. The Radiant code for detecting the 13~activities in the smart healthcare setup translated to 13 Siddhi apps, each between 35 and 50 lines of code, totaling to 545 lines. Based on these numbers, we can observe that Radiant introduces an abstraction/reduction of the required code base by factors 2.4--4.25.} On average, the file size of one generated Siddhi application to be executed on the CEP platform is 5.3\,KB, indicating a relatively small disk memory footprint (Objective~2). Besides the relevant pattern in the sensor data, this also includes all the technical configurations to connect to the event sources and sinks (e.g., the MQTT message broker). As indicated in Section~\ref{sec:rel-ml}, a Siddhi file generated from Radiant is smaller than TinyML-models fine-tuned for activity detection in IoT deployments.

 In contrast to related approaches only working in offline analysis settings, the underlying CEP engine (part of the WSO2 stream processor~\cite{fremantle2015reference}) is specifically designed to process large streams of event data and detect event patterns at runtime, allowing us to provide online feedback regarding the detected activities. The results of our experiments, the small memory and data footprints, and existing benchmarks~\cite{perera2014solving}, indicate that the proposed software architecture (cf.~Section~\ref{sec:runtime}) may serve as an extensible and scalable solution to process IoT data from arbitrary sensors (Objective~4) and to emit process-level events for detected activity executions beyond small scale IoT deployments (cf.~RQ2 and Objective~3). By following best practices from software engineering and architecture, e.g., defining the granularity of one CEP app to match size of one activity, we achieve high cohesion and low coupling in the system, which facilitate extensibility and fault-tolerance~\cite{martin2017clean}. Note that we do not expect the activity detections to be used for real-time control of CPS or their components, but to analyze process executions and provide feedback at runtime to users (e.g.,~a warning in case the healthcare worker forgot to sanitize the hands before performing a specific treatment activity, indicating non-conformance with the process model~\cite{franceschetti2023proambition}) or to other information systems for further analysis or storage.

\subsubsection{Findings: Smart Manufacturing} \label{sec:disc-sm}
The results of the activity detections in the smart manufacturing scenario (cf.~Figure~\ref{fig:fact-ea-ts-f1-act}) show that we can achieve relatively high levels of correct detections when there is only one specific type of activity executed on a production station (e.g., in the sorting machine, oven, and milling machine). Here we are also able to handle variations in sensor patterns quite well via underspecification of patterns and conditions, and alternative patterns (e.g., in the sorting machine). The vacuum gripper robot (VGR) is a central transport entity, which exhibits similar sensor patterns in its movements for different types of activities. Activities can be mostly distinguished based on the x,y,z-target coordinates of the robot, which need to be discretized by the expert first to represent specific locations. A notable station is the high-bay warehouse (HBW) where there are two different types of activities executed (\emph{store} and \emph{unload}). The relevant movement patterns are mostly identical which leads to rather low quality detections. Here we miss a sensor to determine if the buckets being stored or unloaded contain a workpiece or not, which discerns both activities. Moreover, these two activities generally exhibit high amounts of variability as they are executed differently based on the locations of the buckets in the 3x3 storage matrix. Adding more conditions, intermediate and disjunctive patterns might help to increase accuracy here, but will also lead to complex Radiant applications and overfitting, which might decrease the detection quality in the presence of sensor variations again~\cite{seiger5165943online}. Alternatively, the granularity of the activities can be adjusted to, e.g., consider the storage and unloading of an item in each slot of the warehouse as a distinct activity type (e.g.,~store/unload in/from bucket~1). 

Note that in general, the Event Analysis metrics are better for all detections than the Two Set metrics, which also consider the specific points in time when the activity executions happened~\cite{kurz2024activity}. This misalignment can be attributed to the software stack controlling the individual production stations~\cite{seiger2022integrating}. Here we use a layered service-based architecture which introduces some latencies between service executions, logged by the BPM system, and the actual executions in the IoT system that manifest themselves as changes in its sensors and actuators. These latencies might be unintentional (e.g.,~due to network communication) or intentional (e.g.,~the services might perform some calculations or data processing before starting to manipulate the sensors and actuators).

\subsubsection{Findings: Smart Healthcare} \label{sec:disc-sh}
The results of the activity detection in the healthcare scenario (cf.~Figure~\ref{fig:hc-ea-ts-f1-act}) show a lower quality of detections. As described in Section~\ref{sec:setup-healthcare}, the activities in this scenario are almost completely manual and strongly affected by variations in how they are executed, and in the underlying IoT sensor data, which cannot be completely handled by discretizations and alternative detection patterns. While we achieve very good detections for the \emph{Sanitize hands} activity relying on sensor patterns in the scale and distance sensor in front of it (cf.~Figure~\ref{fig:dashboard}), activities related to the interaction with the donor (e.g., apply/remove tourniquet, insert needle, monitor patient) are harder to detect and distinguish from each other as their executions are quite similar and rely on the same sensors and similar patterns (cf.~simultaneous detection of low-level patterns for activities in Figure~\ref{fig:dashboard}). For these activities, we currently consider using cameras in combination with the existing sensors for disambiguation~\cite{franceschetti2025proambition}. Moreover, the starting and stopping of the blood drawing machine rely on the same button to be pressed, thus we cannot distinguish them from each other purely on the sensor data. Currently, we also investigate considering the process context (e.g., which activities happened before) for further disambiguation. Moreover, similar to the event log serving as ground truth for the smart manufacturing settings, we also observe timeshifts when aligning the start and end of activities for the Two Set metrics in the smart healthcare setting, which leads to a decreased performance compared to Event Analysis. These shifts can be attributed to the creation of the ground truth event log, which is based on manual monitoring and tracking by an external observer of the executed processes. Here, the detection of start and end patterns from the IoT sensor data might not always be aligned with the manual tracking. 

\subsubsection{Comparison with Related Approaches}

We are aware of three related approaches that follow the same functional objective in proposing event abstraction mechanisms to detect activity executions in similar IoT settings and based on similar IoT data. In~\cite{fi15020077}, the authors propose a method for domain experts to identify activity executions including the relevant change patterns in the IoT data, and to then manually annotate the data with process level events (i.e., start and end of an activity). This approach relies entirely on the domain expert to manually analyze a given dataset for activity executions (Objective~1), which is not feasible in settings with a high number of sensors and activities that have to be identified and labeled in an unknown dataset.

In~\cite{seiger2025case}, the authors propose to use temporal case-based reasoning for the classification of unknown IoT data according to a known set of activity--IoT data correlations (cases). This approach requires the domain experts to specify similarity measures for the sensors to be used for the matching with the cases. The authors demonstrate that activity detections in the smart factory scenario are possible and feasible (average F1 score of~0.837). While these numbers are an improvement over our experiments, the approach assumes that the IoT data to be classified has already been pre-segmented according to its start and end, and it only works in offline analysis (Objective~2). The underlying models in the case base are more complex than the Radiant applications, but still comprehensible (Objective~1) and relatively light-weight (Objective~3). The approach is applicable to arbitrary sensor data (Objective~4).

In~\cite{garcia2025semi}, the authors propose a sensor data analysis pipeline consisting of clustering and sequence mapping to transform symbolic sequences representing the activity executions into automata, which are able detect the occurrences of the encoded sequences. In the last step, the authors rely on the domain expert to confirm or manually adjust the identified clusters of sensor data and to provide a label that associates the data with an executed activity. Thus, the domain expert is still involved and high F1 scores can be achieved via the manual adjustments. However, the understandability is decreased (Objective~1) due to the clustering and encoding of the multi-variate timeseries of sensor values. The approach is more data and processing intensive (Objective~2) and not capable of online event abstraction (Objective~3).

\subsubsection{Limitations}
Radiant supports domain experts with specifying patterns in sensor data that can be used to detect the execution of process activities. We do not provide any guidance on how to setup the sensor-based activity detection system including which sensors to use or how to combine them to derive process-level events. These are the responsibilities of the IoT engineer and domain expert; our goal is to enable the domain expert to formalize the sensor patterns for automated processing. The sensor patterns and conditions that are currently supported by Radiant are sufficient to cover activity detections in our laboratory setups (cf.~Sections~\ref{sec:setup-manufacturing} and~\ref{sec:setup-healthcare}) as discussed with the involved domain experts. We acknowledge that these environments are simplified setups to simulate real-word IoT systems and leave evaluations in larger scale, more realistic settings to future work. These might for example reveal limitations of the usability of the DSL and performance of the CEP engine when working with complex, high-velocity IoT sensor data, which would require additional abstractions, hierarchical patterns, and pre-processing before patterns can be specified using Radiant. \textcolor{black}{As a specific threat to validity we emphasize again that a formal validation of Radiant's usability has not been conducted so far. Only through this study, which is part of future work, we can evaluate that Radiant does not only support domain experts with specifying sensors patterns for process event abstraction, but that it also does so in an efficient, learnable and usable manner.}

Another specific limitation of the pattern-based approach is that we currently analyze activity executions in isolation, i.e., we do not support multiple concurrent activities to be executed by or across the same stations involving the same sensors at a time, which may lead to a superimposition of sensor data. Furthermore, we experienced some potential issues regarding the conjunction of multiple change conditions in one pattern, which--depending on the sampling frequencies of the sensor data--may also be represented as two subsequent patterns when the changes between two subsequent events from the sensors considered in the change patterns do not happen at the same time. Finally, we did not consider more advanced patterns from the domain of CEP (e.g., stateful patterns, time or count-based aggregations~\cite{etzion2010event}), or relationships among the individual patterns (cf.~Allen's Interval Algebra~\cite{allen1983maintaining,barricelli_visual_2017}) because they did not emerge as requirements in the presented IoT scenarios and also may refer to subsequent process mining analyses, rather than event abstraction. 

In general, our proposed approach is not intended to replace process mining to analyze process and activity executions, but rather to enable it by bringing the IoT sensor data to an appropriate abstraction level where subsequent process mining can provide more reasonable insights. Process mining could already be applied to analyze the low-level IoT sensor data from the sources (e.g.,~to discover processes). However, due to the very fine-grained IoT-based measurements, the complexity of the resulting processes models and process representations is very likely to be too high for analysts to understand and efficiently analyze~\cite{brzychczy2025process}. Finding the correct abstractions and granularity levels in this context~\cite{zerbato2021granularity} is the responsibility of the domain expert using Radiant. Note that while we are emitting process level events (in XES format~\cite{gunther2014xes}) to enable traditional \emph{event-centric} process mining, our approach can also be adapted to \emph{object-centric} event logs~\cite{van2019object} following the models proposed in~\cite{bertrand2023nice} and~\cite{bertrand2025objectcentriccoremetamodeliotenhanced}, which support IoT sensors as data source to be abstracted to process events and correlated with objects.

\section{Conclusion and Future Work} \label{sec:conclusion}
In this work we presented the domain-specific language Radiant, implemented based on the Langium framework, for detecting process activity executions from IoT sensor data. \textcolor{black}{Radiant allows domain experts to specify event abstractions patterns in sensor data that indicate start, end, and intermediate points within the execution of an automated or manual activity.} The Radiant applications are translated to CEP applications and deployed to the corresponding runtime system according to your proposed software architecture, which enables online pattern detection from the sensor streams and subsequent streaming process mining. In contrast to machine learning-based approaches, Radiant applications are light-weight by focusing on sensor patterns not requiring existing data, understandable and easily modifiable by making the sensor patterns explicit, and capable of online activity detection from sensor streams enabled by the software architecture. Domain experts \textcolor{black}{without programming skills} created several Radiant applications to monitor activity executions in smart manufacturing and smart healthcare processes using a variety and number of sensors. The evaluation results show that the CEP apps generated from the Radiant applications can be effectively used to detect process activity executions at scale. Variations in the sensor data and activity execution have a strong influence on the detection quality, which needs to be anticipated by domain expert in the pattern definitions. The results also revealed improvement potential by integrating more advanced concepts into the DSL, considering the process context, and adding sensors to the IoT system.

In future work, we will move the experiments and analysis of patterns to be integrated into Radiant to more complex, real world scenarios and IoT settings. This will also include a more comprehensive user study with domain experts to evaluate the usability and efficiency of Radiant, \textcolor{black}{which has not been evaluated in terms of usability so far}. Furthermore, we will use the results of the activity detections in subsequent online process mining analyses to check for process conformance at runtime and resolve potential ambiguities based on the process context. This way we will be able to provide online feedback and suggestions about process conformance to the end-users. We will also investigate the combination of ML-based approaches together with pattern specification in Radiant resulting in a hybrid activity detection system. We envision that having a robust way of detecting and representing a process activity based on sensor data will facilitate the development of digital twins of business processes enabled by the IoT~\cite{fornari2024digital}.

\section*{Acknowledgments}

This work has received funding from the Swiss National Science Foundation under Grant No.~IZSTZ0\_208497 (\textit{Pro\-AmbitIon} project). This work has been supported by the Internet of Processes and Things (IoPT) community through discussions, datasets, and feedback.

\bibliographystyle{elsarticle-num} 
\bibliography{refs}

\end{document}